\documentclass[aps,pre,twocolumn,superscriptaddress,amsmath,amssymb,longbibliography]{revtex4}

\usepackage{graphicx}

\usepackage{dcolumn}

\usepackage{enumitem,kantlipsum}

\numberwithin{equation}{section}

\usepackage{bm}
\usepackage[normalem]{ulem}
\usepackage{color}
\usepackage{gensymb}

\usepackage{epstopdf}

\usepackage{xcolor}
\usepackage{color}

\definecolor{green}{rgb}{0,0.5,0}

\begin{document}
 
\title{Steady states and phase transitions in heterogeneous asymmetric exclusion processes}
\author{Atri Goswami}\email{goswami.atri@gmail.com}
\affiliation{Gurudas College, 1/1, Suren Sarkar Road, Jewish Graveyard, Phool Bagan, Narkeldanga, Kolkata, West Bengal 700054}
\author{Mainak Chatterjee}\email{mainak216@gmail.com}
\affiliation{Barasat Government College,
10, KNC Road, Gupta Colony, Barasat, Kolkata 700124,
West Bengal, India}
\author{Sudip Mukherjee}\email{sudip.bat@gmail.com, sudip.mukherjee@saha.ac.in}
\affiliation{Barasat Government College,
10, KNC Road, Gupta Colony, Barasat, Kolkata 700124,
West Bengal, India}

\begin{abstract}

We study the nonequilibrium steady states in  totally asymmetric exclusion processes (TASEP) with open boundary conditions having spatially inhomogeneous hopping rates. Considering smoothly varying hopping rates, we show that the steady states are in general classified by the steady state currents in direct analogy with open TASEPs having uniform hopping rates.  
We calculate the steady state bulk density profiles, which are now spatially nonuniform. We also obtain the phase diagrams in the plane of the control parameters, which though have phase boundaries that are in general curved lines, have the same topology as their counterparts for conventional open TASEPs, independent of the form of the hopping rate functions. This reveals a type of universality, not encountered in critical phenomena.  Surprisingly and in contrast to the phase transitions in an open TASEP with uniform hopping,  our studies on the phase transitions in the model reveal that all the three transitions are {\em first order} in nature. { We also demonstrate that this model admits delocalised domain walls (DDWs) on the phase boundaries demarcating the generalised low and high density phases in this model. However, in contrast to the DDWs observed in an open TASEP with uniform hopping, the envelopes of the DDWs in the present model are generally curved lines.}
 
\end{abstract}

\maketitle

 \section{Introduction}
 
 
 Many natural systems are driven by some external fields or
are made of self-propelled particles. In the long time limit, these systems evolve into stationary
states which carry  steady currents, which are hallmarks of nonequilibrium systems. Such states are characterised
by a constant gain or loss of energy, which distinguishes
them from systems in thermal equilibrium. Examples of such driven systems range from live cell biological systems like ribosomes moving along mRNA or motor
molecules “walking” along molecular tracks known as microtubules to ions diffusing
along narrow channels, or even vehicles traveling along roads. In order to elucidate the nature of such nonequilibrium
steady states and in the absence of a general theoretical framework, it is useful to study purpose-built simple models. To this end, a variety of driven lattice gas models have been
introduced and studied extensively~\cite{basic}.

In this work, we focus on driven
one-dimensional (1D) models with open boundaries, where particles preferentially
move in one direction. In particular, we work on  the totally asymmetric
simple exclusion process (TASEP), that has become one of the
paradigms of nonequilibrium physics in low-dimensional systems (see, e.g., Ref.~\cite{tasep-rev}
for reviews). In this model identical particles   hop
unidirectionally and with a uniform rate along a 1D lattice~\cite{krug}. The hopping movement is subject to exclusion, i.e., when the target site is empty, since a given site can accommodate maximum one particle. Particles enter the system at one side at a specified rate $\alpha$, and leave the system through the other end at a given rate $\beta$; $\alpha$ and $\beta$ are the two control parameters of TASEP. It is known that the steady states of TASEPs with open boundaries are rather sensitive to the boundary conditions: by varying the boundary conditions, i.e., by varying $\alpha,\beta$, the steady states of open TASEPs can be varied, resulting into {\em boundary-induced nonequilibrium phase transitions}. These are genuine nonequilibrium effects, since equilibrium systems are usually insensitive to precise boundary conditions.

In the original TASEP model, the hopping rate in the bulk is assumed to be a constant (of unit value), which is of course an idealisation. In real life examples it is generally expected to have nonuniformity along the bulk of the TASEP channel leading to nonuniform hopping rates. For instance, mRNA in cells are known to have pause sites, where the effective hopping rates are lower~\cite{pause}. This is a potentially important issue even in urban transport, where the speeds of vehicles (which is the analogue of the hopping rates here) depend sensitively on the bottlenecks along the roads~\cite{transport}. Such spatially varying hopping rates can either be smoothly varying along the TASEP lanes, or be random quenched disorders with given distribution. We focus here on the case with smoothly varying hopping rates, for which the generic nature of the steady states in TASEPs are still not known. There have been some studies on quenched heterogeneous TASEP; see, e.g., Refs.~\cite{qtasep1,qtasep} for previous studies on different
aspects of heterogeneous TASEP. Recently, how the steady states of TASEPs with periodic boundary conditions are affected by smoothly varying hopping rates are studied~\cite{prr}; see also Ref.~\cite{astik-prr} for a study on periodic TASEP with random quenched disordered hopping rates. A type of universality has been uncovered, showing the topological equivalence of the phase diagrams independent of the precise form of the space dependence of the hopping rates.  More recently, an inhomogeneous $\ell$-TASEP with open boundaries has been proposed in the context of ribosome movements along messenger RNA strands~\cite{erdmann}. Using a hydrodynamic approach~\cite{erdmann}, this study reveals complex natures of the steady states and the phase transitions. In the present work, we revisit the problem of open TASEPs with spatially non-uniform hopping rates, which corresponds to the $\ell=1$ limit of the model studied in Ref.~\cite{erdmann}. 
Following Ref.~\cite{prr}, we set up the analytical mean-field theory (MFT) framework to calculate the steady state density profiles for generic smoothly varying hopping rates. We illustrate the theoretical predictions by calculating the density profiles for a few representative examples of spatially varying hopping rates in Monte-Carlo simulation (MCS) studies.  The MFT that we develop  complements the hydrodynamics-based approach developed in Ref.~\cite{erdmann}, and helps to understand the applicability and limitations of MFT in heterogeneous TASEPs, an important theoretical and technical issue given MFT's prevalence and general popularity as a theoretical tool to study exclusion processes. Within this MFT, by defining an order parameter reminiscent of the same in open TASEPs but independent of the forms of the space dependent hopping rates, we are   able to clearly connect the quantitative difference between the phase diagrams obtained here with those for open TASEPs with uniform hopping.  
We further study the nonequilibrium phase transitions in the model. Our MCS studies show that {\em all} the phase transitions are first order or discontinuous in nature, a surprising and unexpected outcome from this work, that is easily explained within the MFT we develop here. This is in contrast to an open TASEP with uniform hopping. Our extensive MCS studies on the domain walls reveal their delocalised nature, qualitatively similar to those found in an open uniform TASEP. Nonetheless, our studies point out how the effects of the nonuniform hopping get visible in the form of the envelope of the moving domain wall, which is now a nonlinear function of position along the TASEP channel. The rest of the article is organised as follows. In Sec.~\ref{model} we define and construct our model. Next, in Sec.~\ref{monte} we discuss the algorithm of the MCS study of the model to numerically calculate the steady state densities. Then in Sec.~\ref{mean}, we set up the MFT, and solve it to obtain the steady state densities for smoothly varying hopping rates, and compare with our MCS results. In Sec.~\ref{phase}, we present the phase diagrams of the model.  Then in Sec.~\ref{phase-trans}, we discuss the phase transitions in the model. We summarise our results in Sec.~\ref{summ}.
 
 \section{Model}\label{model}
 
\begin{figure}[htb]
 \includegraphics[width=1.1\columnwidth]{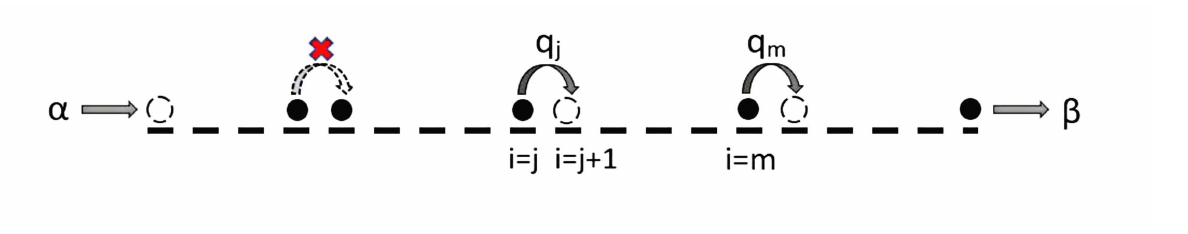}
  \caption{ Schematic model diagram. Broken line represents the TASEP lattice. Particles enter and exit at rates $\alpha$ and $\beta$, respectively, and hop from left to right, subject to exclusion.}\label{modelfig}
 \end{figure}

 
 The model consists of a 1D lattice of size $L$. The particles enter through the left end at rate $\alpha$, hop unidirectionally from the left to the right, all subject to exclusion, i.e., a single site can accommodate maximum one particle at a time, and finally leave the system at a rate $\beta$. Labelling each site by an index $i$ that runs from 1 to $L$, the hopping rate at site $i$ is given by $q_i\leq 1$; see Fig.~\ref{modelfig} for a schematic model diagram. 
 
 A microscopic configuration of the model is characterised by a distribution of identical particles on the lattice, i.e., by configurations ${\cal C} = \{n_{i=1,. . .,L}\}$, where each of the occupation numbers $n_i$ is equal to either zero (vacancy) or one (particle), as it should be in a model with exclusion. Physically,
 a hard core repulsion between the particles is imposed, resulting into prohibition of
 a double or higher occupancy of sites in the model. The full state space then consists of $2^L$
configurations. The following elementary processes fully define the microscopic dynamical update rules of this model:

\noindent (a) At any site $i = 1 , . . . , L-1$ a particle can jump to site $i+ 1$ if unoccupied with a rate $q_i\leq 1$.

\noindent (b) At the site $i = 1$ a particle can enter the lattice with rate $\alpha q(1)$
only if it is unoccupied; and

\noindent (c) At the site $i = N$ a particle can leave the lattice with rate
$\beta q(L)$ when it is occupied.

In general, $q_i\neq q_j$ for $i\neq j$. Processes (a)-(c) formally define a TASEP with open boundary conditions. If all of $q_i=1$ for all $i$ identically, then this model reduces to the conventional TASEP with open boundary conditions~\cite{tasep-rev}. We consider some specified choices of $q_i$ that depends explicitly on $i$, and study their effects on the nonequilibrium steady states of the model.  Recall that the steady states of an open TASEP with $\alpha$ and $\beta$ as the entry and exit rates, and a uniform hopping rate are characterised by the mean bulk density $\rho_T$: For $\alpha<\beta$ and $\alpha<1/2$, one has $\rho_T=\alpha$ giving the low density (LD) phase, for $\beta<\alpha$ and $\beta<1/2$, one has $\rho_T=1-\beta$ giving the high density (HD) phase, and for $\alpha,\,\beta>1/2$, one has $\rho_T=1/2$ giving the maximal current (MC) phase. This immediately gives the phase boundary in the $\alpha-\beta$ plane~\cite{tasep-rev}. The principal aim of the present study is to find the phases, phase boundaries and the nature of the associated phase transitions, and the principles behind obtaining them when the hopping rate is not constant, but spatially smoothly varying.

 \section{Steady-state densities}
 
 We are interested to calculate the density profiles in the steady states. To this end, we set up MFT which can be solved analytically. We supplement the MFT results by extensive Monte-Carlo simulations (MCS).
 
 \subsection{Monte-Carlo simulations}\label{monte}
 
  We consider a lattice of $L$ sites, labelled by an index $i$ with $i\in [1,L]$.
 Let $n_i(t)$, which is either 0 or 1, be the occupation at site $i$ at time $t$.  We perform MCS studies of the model subject to the update rules (a)-(c) described above in Sec.~\ref{model} by using a random sequential updating scheme. The particles enter the system through the left most site ($i=1$) at a fixed rate $\alpha$, subject to exclusion, i.e., if $n_1=0$. After hopping through the system from $i=1$ to $L$, subject to exclusion, the particles exit the system from $i=L$ at a fixed rate $\beta$. Here, $\alpha$ and $\beta$ are the two simulation parameters, which are varied to produce different steady states. After reaching the steady states, the density profiles are calculated and temporal averages are performed. This produces time-averaged, space-dependent density profiles, given by $\langle n_i\rangle$,which are parametrised by $\alpha$ and $\beta$; here $\langle...\rangle$ implies temporal averages over steady states. The simulations have been performed with  $L=10000$ up to $10^7$ Monte-Carlo steps. 
 { Lastly, all the measurements are made in the steady states, which are reached by the system after spending certain transient times. In an open TASEP, a steady state is easily ascertained by observing the spatio-temporal constancy of the average density $\langle n_i(t)\rangle$ (excluding the domain walls) in the bulk of the system. In the present problem, such a way to confirm the steady state fails due to the (expected) spatially varying steady state density in the bulk. Instead, we use the constancy of the current $J$ in the steady state, a condition that holds both in the present study and also for a uniform open TASEP. In our MCS studies, all our measurements are done only after this condition is satisfied.} 
 
 \subsection{Mean-field theory}\label{mean}
 
 
 The dynamics of TASEP is formally given by rate equations for every site which are {\em not} closed. In MFT approximation, we neglect correlation effects and replace the average of product of densities by the product of average of densities~\cite{blythe}. While this is an approximation, this has worked with high degree of accuracy in the original TASEP problem and its many variants (see, e.g., Refs.~\cite{erwin-lk,niladri1,tirtha1} as representative examples); we use MFT here as a guideline in our analysis below. The dynamical equation for $n_i(t)$ is given by
 \begin{equation}
  \frac{\partial n_i}{\partial t} = q_in_{i-1}(1-n_i) - q_{i+1}n_i(1-n_{i+1}),\label{basiceq-0}
 \end{equation}
  for a site $i$ in the bulk. Clearly, Eq.~(\ref{basiceq-0}) is invariant under the transformation $n_i(t) \rightarrow 1-n_{L-i}(t)$ together with $q_i \rightarrow q_{L-i}$ and $\alpha \rightarrow \beta$, which is the {\em particle-hole symmetry} of this model~\cite{erwin-lk}. 
 
 To proceed further in the MFT approximation, we label the sites by $x=i/L$ and take $L\rightarrow \infty$, which makes $x$ a continuous variable between 0 and 1: $x\in [0,1]$. In this parametrisation, the hopping rate function is given by $0<q(x)\leq 1$, that is assumed to vary slowly in $x$. We define a lattice constant $\varepsilon \equiv L_0/L$, where $L_0$ is the geometric length of the system. To simplify notation, we fix the total length  $L_0$ to unity without any loss of generality. In the thermodynamic limit $L\rightarrow \infty$, $\varepsilon\rightarrow 0$ is a small parameter. Further, we define $\rho(x)=\langle n_i\rangle$ as the steady state density at $x$. In the steady state, we expand the different terms on rhs of (\ref{basiceq-0}) in a Taylor series in powers of $\varepsilon$. We get
 \begin{eqnarray}
  \rho(x\pm\varepsilon)&=& \rho(x)\pm \varepsilon\partial_x\rho(x) \nonumber \\&&+\frac{\varepsilon^2}{2}\partial_x^2 \rho(x) + {\cal O} (\varepsilon^3),\\
  q(x\pm \varepsilon)&=& q(x) \pm \varepsilon \partial_x q(x) \nonumber \\&&+ \frac{\varepsilon^2}{2}\partial_x^2 q(x) + {\cal O} (\varepsilon^3).
 \end{eqnarray}
Substituting the above and retaining up to ${\cal O}(\varepsilon)$, we get 
 
 \begin{equation}
  \frac{\partial \rho}{\partial t} = - \varepsilon\frac{\partial}{\partial x}\left[q(x)\rho(x)(1-\rho(x))\right] { + {\cal O} (\varepsilon^2)} ,\label{mft1}
 \end{equation}
neglecting terms higher order in $\varepsilon$. Equation~(\ref{mft1}) allows us to extract a bulk current $J$ given by
\begin{equation}
 J=q(x)\rho(x)[1-\rho(x)] { + {\cal O}(\varepsilon)},\label{curr0}
\end{equation}
which must be a constant independent of $x$ in a given steady state. That Eq.~(\ref{mft1}) has the form of an equation of continuity is no surprise - this is because away from the boundaries in the bulk of the TASEP, particles only hop from left to right, subject to exclusion, which keeps the particle number locally conserved. In the continuum limit, $\varepsilon\rightarrow 0^+$ and hence the average current is
\begin{equation}
 J=q(x)\rho(x)[1-\rho(x)],\label{curr}
\end{equation}
valid when $\rho(x)$ and $q(x)$ are sufficiently smooth.  It is evident that the MFT equations (\ref{mft1})-(\ref{curr}) are invariant under the particle-hole symmetry discussed above. Due to this property it is enough to restrict the analysis to the LD and MC phase density profiles; the HD phase density can be constructed from the LD phase density by using the particle-hole symmetry.  Notice that this MFT does not capture any time-dependent or dynamical information, unlike hydrodynamic approaches. Nonetheless, considering the inherent simplicity of our MFT and the general popularity of MFT as a theoretical tool to study TASEPs, it is useful to study the steady states of this model by MFT, which serves as a good benchmark of the success, applicability and limitations of the mean field approaches vis-\'a-vis other approaches like hydrodynamic methods.

\paragraph{General solutions of the density in MFT:-}  We now derive the generic steady state density profiles and delineate the phases by using the MFT equation (\ref{curr}). {In this study, we closely follow the method outlined in Ref.~\cite{prr}, that was subsequently extended to and refined for interacting systems~\cite{arvind1,arvind2}.} As argued below, these solutions holds for any smoothly varying $q(x)$. Equation~(\ref{curr}) is a quadratic equation in $\rho(x)$.
In Eq.~(\ref{curr}) since $J$ is a constant and $q(x)$ has an explicit $x$-dependence, $\rho(x)$ must be $x$-dependent, such that the product of the various factors on the rhs of (\ref{curr}), all of which are individually $x$-dependent, produces an $x$-independent result $J$. Equation~(\ref{curr}) has two spatially nonuniform solutions $\rho_+(x)$ and $\rho_-(x)$ for a given $J$:
\begin{eqnarray}
 && \rho_+(x)=\frac{1}{2}\left[1+\sqrt{1-\frac{4J}{q(x)}}\right]>\frac{1}{2},\label{rho+}\\
 &&\rho_-(x)=\frac{1}{2}\left[1-\sqrt{1-\frac{4J}{q(x)}}\right]<\frac{1}{2},\label{rho-}
\end{eqnarray}
for any $x$. Evidently, both $\rho_+(x)$ and $\rho_-(x)$ are   continuous functions of $x$,   as long as $q(x)$ itself is a continuous function of $x$. Further, $\rho_+(x)>1/2$ everywhere, whereas $\rho_-(x)<1/2$ everywhere. Clearly, if $q(x)$ is a  constant then $\rho(x)$ is also a constant, independent of $x$ (ordinary TASEP with uniform hopping). At this stage $J$ is still unknown. Since $\rho(x)$ is  real (in fact positive definite) everywhere, we must have $1-4J/q(x)\geq 0$ giving an upper bound on $J$:
\begin{equation}
 J\leq q(x)/4.\label{J-ineq}
\end{equation}
 Inequality (\ref{J-ineq}) must hold for {\em all} $x$. Clearly, $J$ has a maximum given by
\begin{equation}
 J_\text{max} = \frac{q_\text{min}}{4},
\end{equation}
for a given $q(x)$; see also Ref.~\cite{krug11} for an analogous result in a disordered exclusion process.
Note that $J_\text{max}$ is the {\em maximum possible} current that can be sustained by the system for all possible choices of the control parameters $\alpha,\,\beta$. However, $J$ may not reach $J_\text{max}$ for any $\alpha,\,\beta$; see below. In the limit of uniform hopping with $q(x)=1$ everywhere, $J_\text{max}=1/4$, corresponding to the MC phase current in the conventional TASEP. We thus note that in the present model steady states with current $J=J_\text{max}$ for a given $q(x)$ should generalise the standard MC phase in open TASEPs with uniform hopping. 

We now systematically derive the conditions for the different phases. To do this, we must calculate $J$ to specify the solutions $\rho_+(x)$ and $\rho_-(x)$ completely. 

Recall that in the LD phase of conventional open TASEPs with uniform hopping, the steady state is described by the incoming current, which in the bulk is given by  $J^T_\text{LD}=\alpha(1-\alpha)< J^T_\text{HD}=\beta(1-\beta)$, the outgoing current. This corresponds to $\rho=\alpha$ as the bulk density in the LD phase. In contrast, in the HD phase, $J^T_\text{HD}<J^T_\text{LD}$, giving $1-\beta$ as the HD phase bulk density; a superscript $T$ refers to an open TASEP with uniform hopping. The MC phase is  associated with the current $J^T_\text{MC}=1/4$ in the bulk. The LD-HD phase boundary is given by the condition $J^T_\text{LD}=J^T_\text{HD}$ giving $\alpha=\beta$; the LD-MC and HD-MC phase boundaries likewise are given by $J^T_\text{LD}=J^T_\text{MC}$ and $J^T_\text{HD}=J^T_\text{MC}$, giving, respectively, $\alpha=1/2$ and $\beta=1/2$ as the phase boundaries. We now generalise this picture by finding out the forms of $J$ in the present problem. 
 
 \paragraph{LD phase:-} We start by noting that the current in the bulk of the TASEP channel is $J = q(x) \rho (x) [1- \rho(x)]$, where $\rho(x)$ is $\rho_+(x)$ or $\rho_-(x)$. Since $\rho(0)=\alpha$, we obtain 
 \begin{equation}
 J_\text{LD}=q(0)\alpha(1-\alpha). \label{jld}
 \end{equation}
  Since, the steady state bulk density in the LD phase is less than 1/2 everywhere, the density profile $\rho_\text{LD}(x)$ in the LD phase is given by
 \begin{equation}
  \rho_\text{LD}(x)=\frac{1}{2}\left[1-\sqrt{1-\frac{4q(0)}{q(x)}\alpha(1-\alpha)}\;\right]<\frac{1}{2}.\label{rhold}
 \end{equation}
  Then we must have $\alpha<1/2$ as in an open TASEP with a uniform hopping rate.
 Equation~(\ref{rhold}) gives the steady state density in the LD phase for a given $q(x)$ and depends on the entry rate $\alpha$, but not on the exit rate $\beta$, as expected in the LD phase. With $q(x)=q(0)=const.$ everywhere, $\rho_\text{LD}(x)=\alpha$, neglecting the other solution $1-\alpha>1/2$ for $\alpha<1/2$ for an open TASEP with a uniform hopping rate.
 
\begin{widetext}

 \begin{figure}[htb]
  \includegraphics[width=8.3cm]{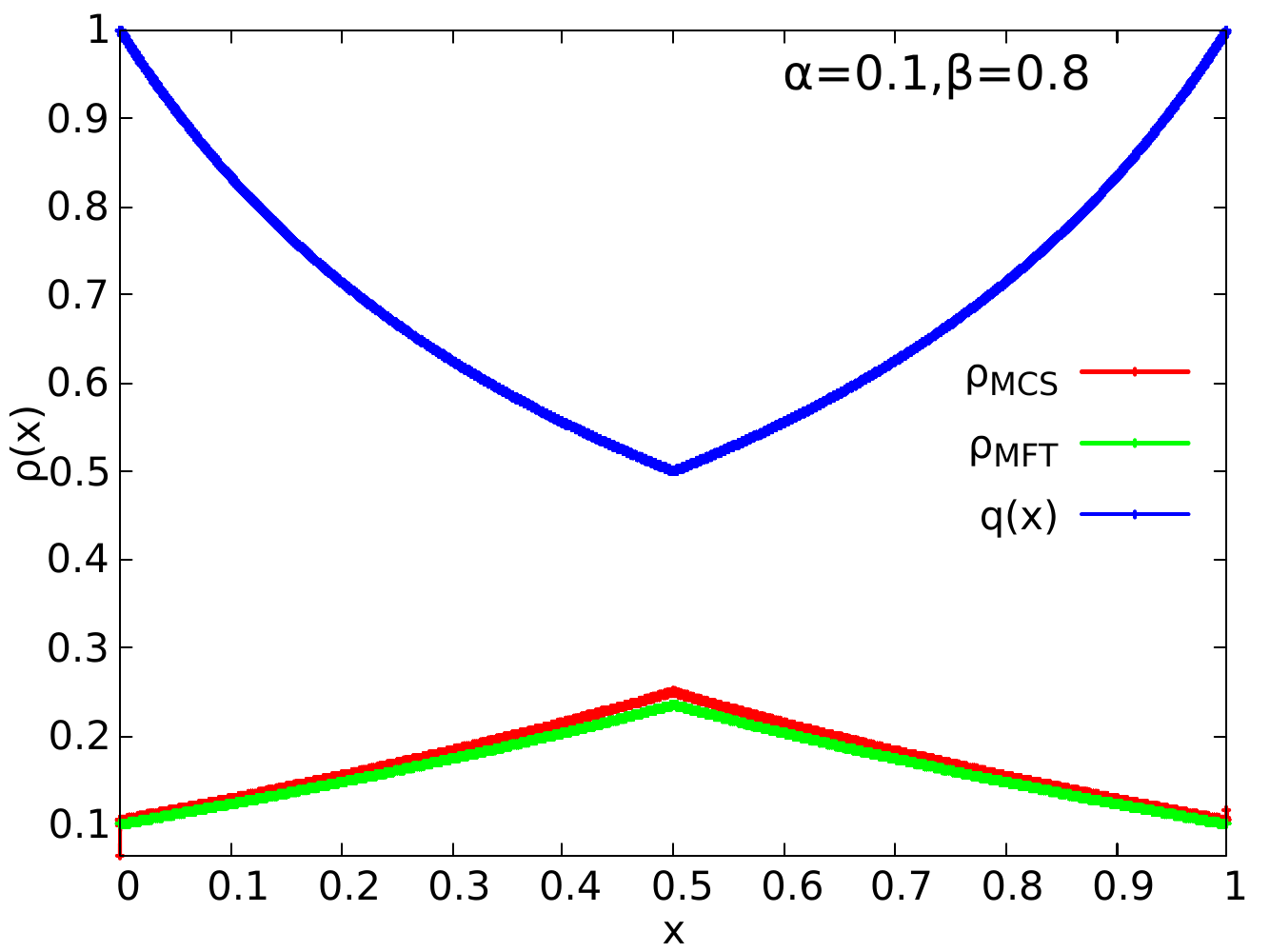}\hfill
  \includegraphics[width=8.3cm]{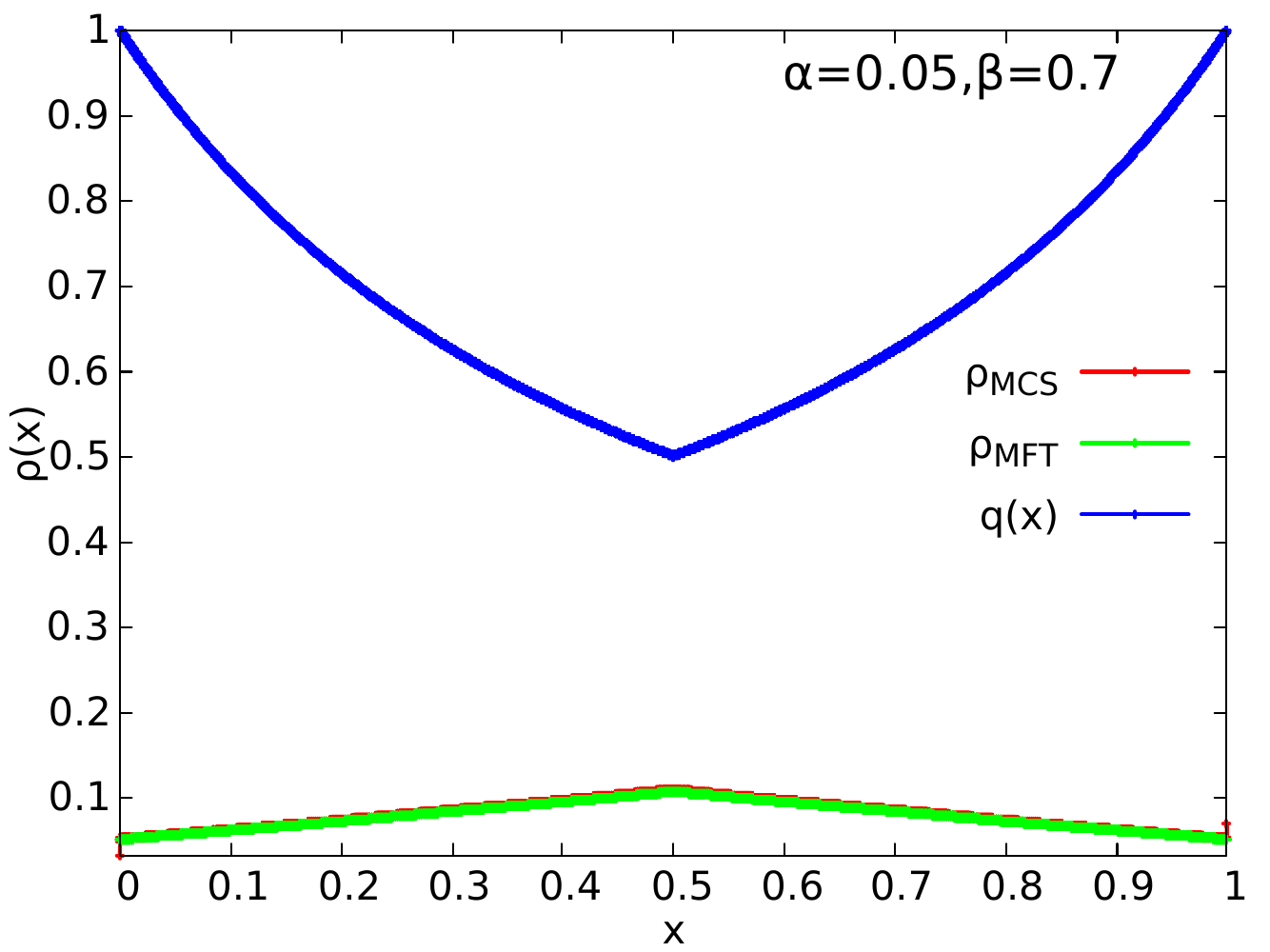}\vskip0.7cm
  \includegraphics[width=8.3cm]{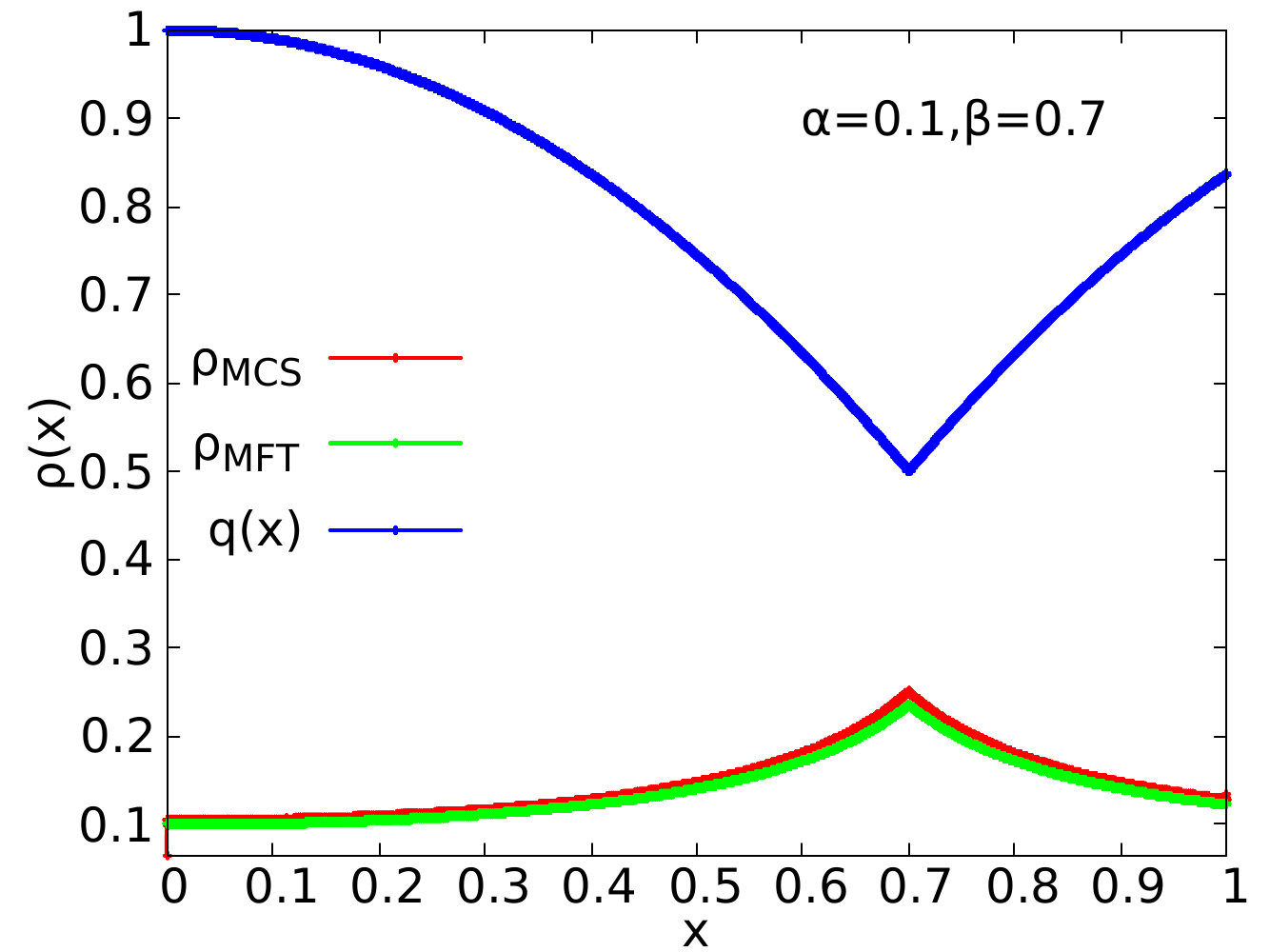}\hfill
  \includegraphics[width=8.3cm]{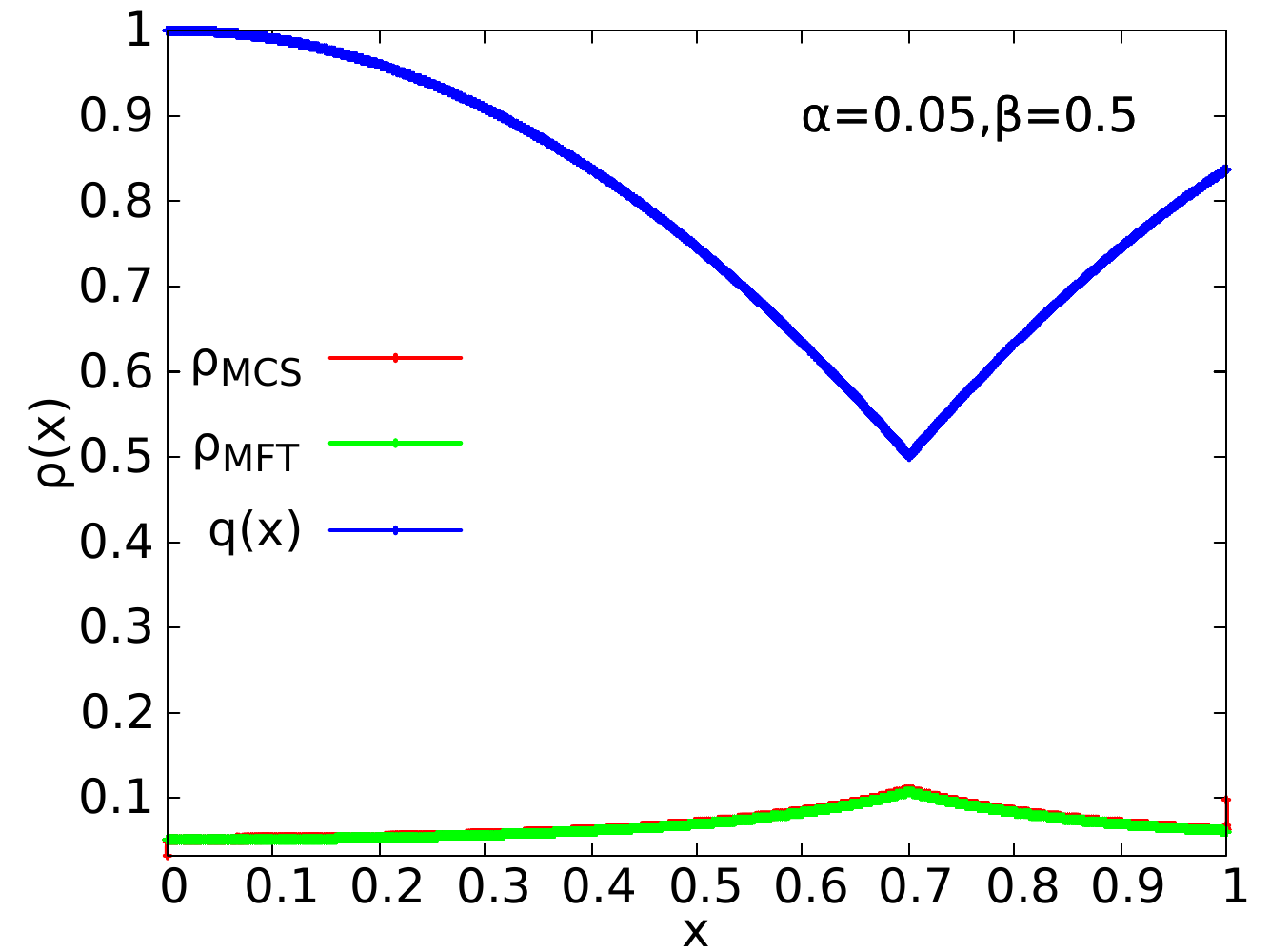}
  \caption{Plots of the steady state density $\rho(x)$ versus $x$ in the LD phase for different choices of the hopping rate functions. (top) $q(x)$ as given in Choice I above for two different sets of values of $\alpha$ and $\beta$, (bottom) $q(x)$ Choice II above for two different sets of $\alpha$ and $\beta$. In each plot, the green line represents the MFT prediction, the red points are from the corresponding MCS study; the blue line represents $q(x)$. Good agreement between the MFT and MCS predictions can be seen (see text).}\label{ld-plot}
 \end{figure}
 
 \end{widetext}

 We have plotted $\rho_\text{LD}(x)$ versus $x$ in Fig.~\ref{ld-plot} for two different and simple choices of the hopping rate function $q(x)$: 
 \begin{eqnarray}
  \text{Choice I:}\;\;q(x)&=&\frac{1}{1+2x},\;\;0\leq x\leq 1/2,\nonumber \\
  &=&\frac{1}{3-2x},1/2\leq x\leq 1,\label{choice1}\\
  \text{Choice II:}\;\;q(x)&=&\frac{1}{2}\left[2-\frac{x^2}{0.49}\right],\;\;0\leq x\leq 0.7,\nonumber \\
  &=& \frac{1}{2}\left[2-\frac{(x-1.4)^2}{0.49}\right],\;\;0.7\leq x\leq 1.\label{choice2}\nonumber \\
 \end{eqnarray}
Clearly, $q(x)$ in Choice I is {\em symmetric} about $x=1/2$, whereas $q(x)$ in Choice II has {\em no} particular symmetry.
  Results on the steady state densities from MFT and MCS studies are plotted together in Fig.~\ref{ld-plot}, which show good agreements between MFT and MCS results. 
 
\paragraph{HD phase:-} The logic we have developed above to obtain $\rho_\text{LD}(x)$ can be used to obtain $\rho_\text{HD}(x)$, the steady state density in the HD phase. Noting that $\rho(1)=1-\beta$, we obtain the HD phase current
\begin{equation}
J_\text{HD}=q(1)\beta(1-\beta).\label{jhd}
\end{equation}
Since, the steady state bulk density everywhere is more than 1/2, the density profile $\rho_\text{HD}(x)$ in the HD phase is given by
 \begin{equation}
  \rho_\text{HD}(x)=\frac{1}{2}\left[1+\sqrt{1-\frac{4q(1)}{q(x)}\beta(1-\beta)}\;\right]>\frac{1}{2}.\label{rhohd}
 \end{equation}
  Therefore, we must have $\beta<1/2$ as for an open TASEP with uniform hopping.
 In contrast to $\rho_\text{LD}(x)$, given by (\ref{rhold}) above, $\rho_\text{HD}(x)$ in (\ref{rhohd}) depends on $q(x)$ and the exit rate $\beta$, but not on $\alpha$, as expected in the HD phase. With $q(x)=q(1)=const.$ everywhere, $\rho_\text{HD}(x)=1-\beta$, neglecting the other solution $\rho_\text{HD}=\beta<1/2$  for an open TASEP with a uniform hopping rate.
 
 In an open TASEP with uniform hopping, $\alpha<1/2$ and $\alpha<\beta$ specify the LD phase, whereas $\beta<\alpha$ and $\beta<1/2$ specify the HD phase. What are the analogous conditions here? These conditions in the present case may be obtained by considering the steady state currents. We recall that the above conditions for the LD and HD phases in an open TASEP with uniform hopping can be recast in terms of the steady state currents as $J^T_\text{LD}<1/4$ and $J^T_\text{LD}<J^T_\text{HD}$ for the LD phase, and $J^T_\text{HD}<1/4$ and $J^T_\text{HD}<J^T_\text{LD}$ for the HD phase. These conditions may be generalised to the present case with non-uniform hopping. The LD phase now exists for
 \begin{equation}
  J_\text{LD}\equiv q(0)\alpha(1-\alpha)<J_\text{HD}\equiv q(1)\beta(1-\beta),\;J_\text{LD}<\frac{q_\text{min}}{4}. \label{ld-cond}
 \end{equation}
 Similarly, for the HD phase to exist we must have
 \begin{equation}
  J_\text{HD}<J_\text{LD},\;J_\text{HD}<\frac{q_\text{min}}{4}.\label{hd-cond}
 \end{equation}

\begin{figure}[htb]
  \includegraphics[width=\columnwidth]{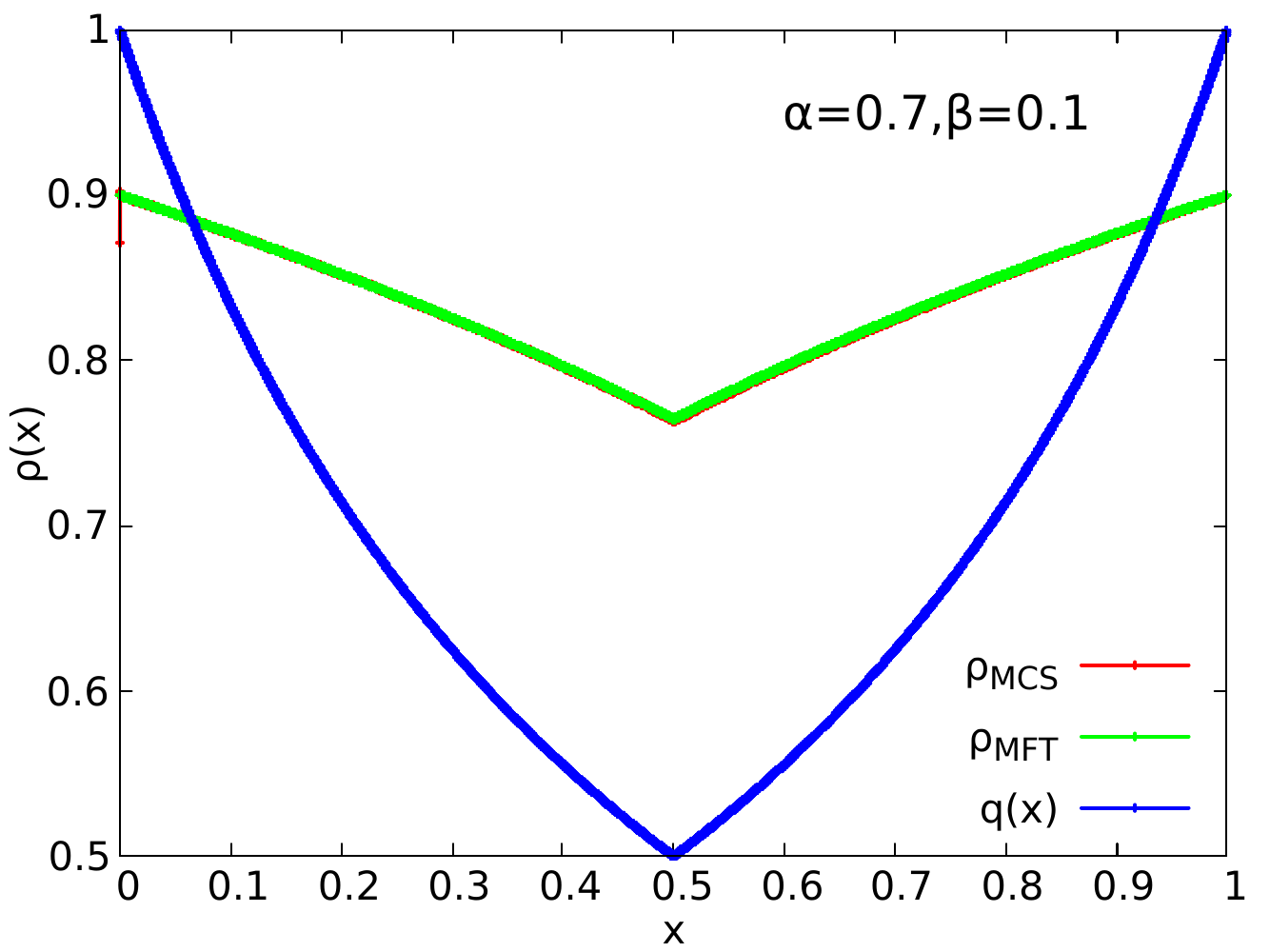}\vskip0.5cm
  \includegraphics[width=\columnwidth]{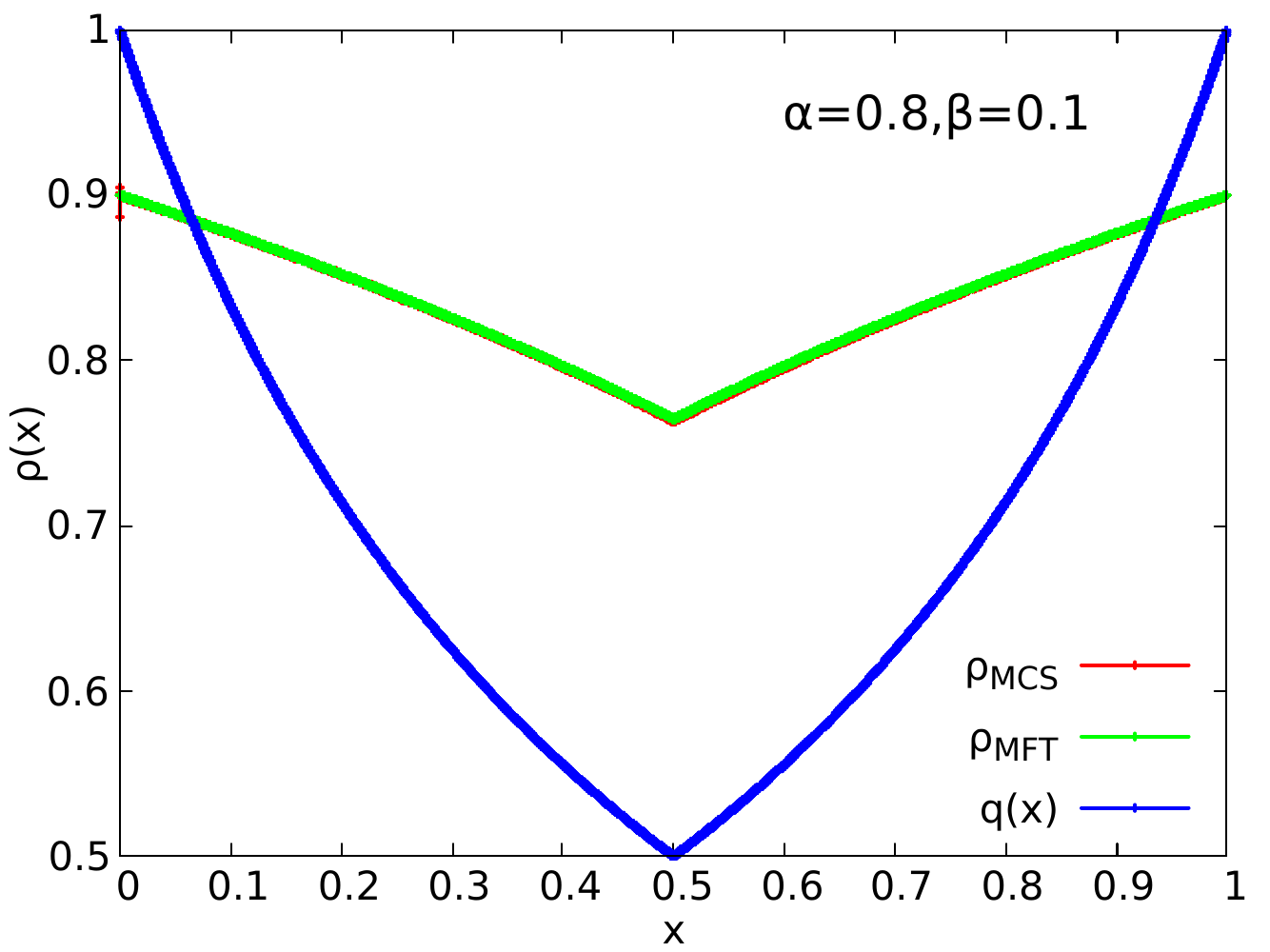}
  \caption{ Plots of the steady state density $\rho(x)$ versus $x$ in the HD phase for the hopping rate function $q(x)$ as given in Choice I above for two different sets of values of $\alpha$ and $\beta$.  These are connected to the corresponding $\rho_\text{LD}(x)$ via the particle-hole symmetry discussed above.  In each plot, the green line represents the MFT prediction, the red points are from the corresponding MCS study; the blue line represents $q(x)$. Good agreement between the MFT and MCS predictions can be seen (see text).}\label{hd-plot}
 \end{figure}

 We have plotted $\rho_\text{HD}(x)$ versus $x$ in Fig.~\ref{hd-plot} for $q(x)$ as defined in Choice I above. The HD phase density plots for $q(x)$ as given in Choice II above, likewise, can be obtained from corresponding plots in the LD phase by using the particle-hole symmetry.
  Results from MFT and MCS studies are plotted together in Fig.~\ref{hd-plot}, which again reveal good agreements between MFT and MCS results.

  At this stage, we note that the MFT density expressions agree with the predictions from the hydrodynamic theory~\cite{erdmann}.

\begin{widetext}

  \begin{figure}[htb]
  \includegraphics[width=8.3cm]{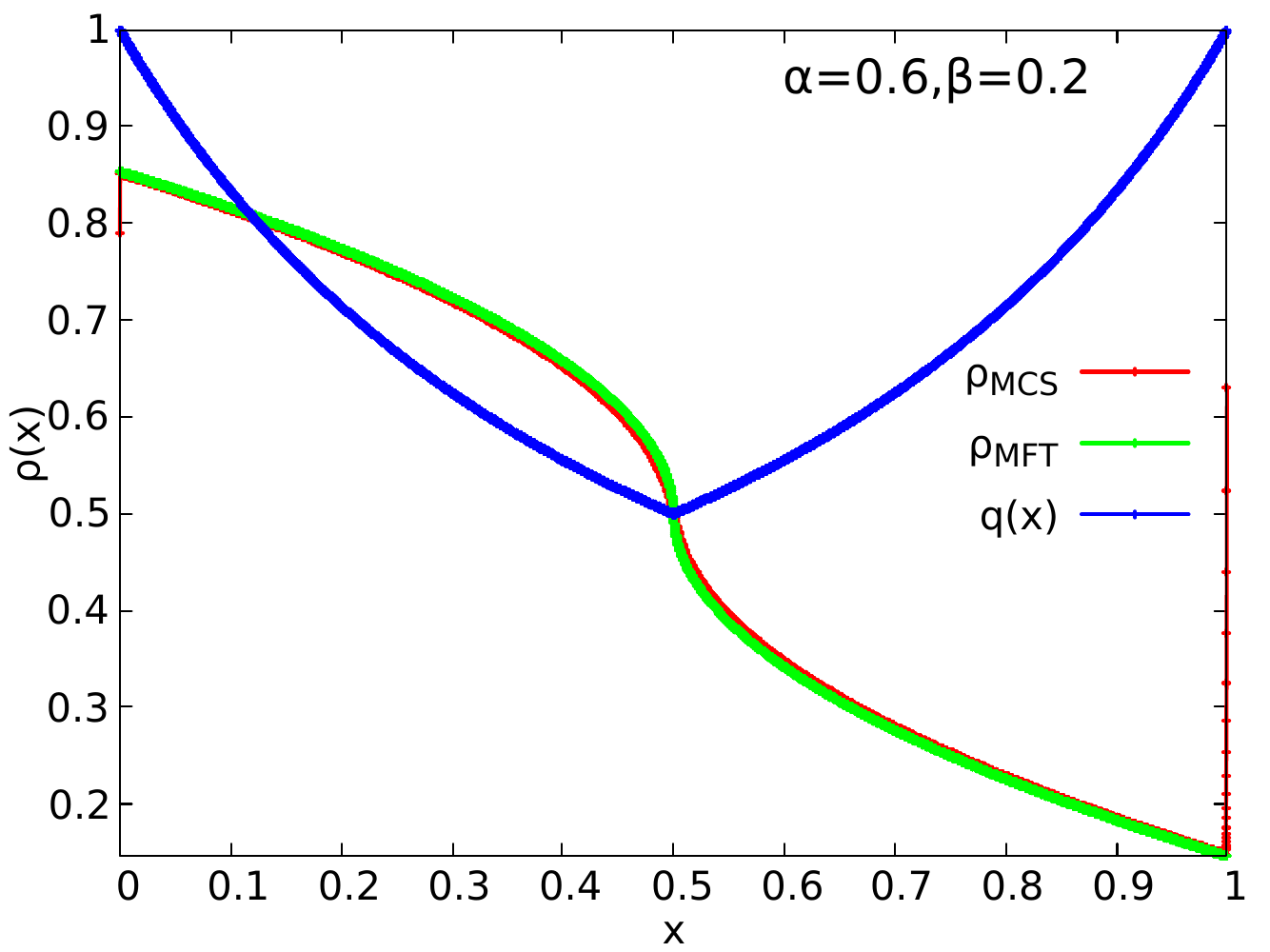}\hfill
  \includegraphics[width=8.3cm]{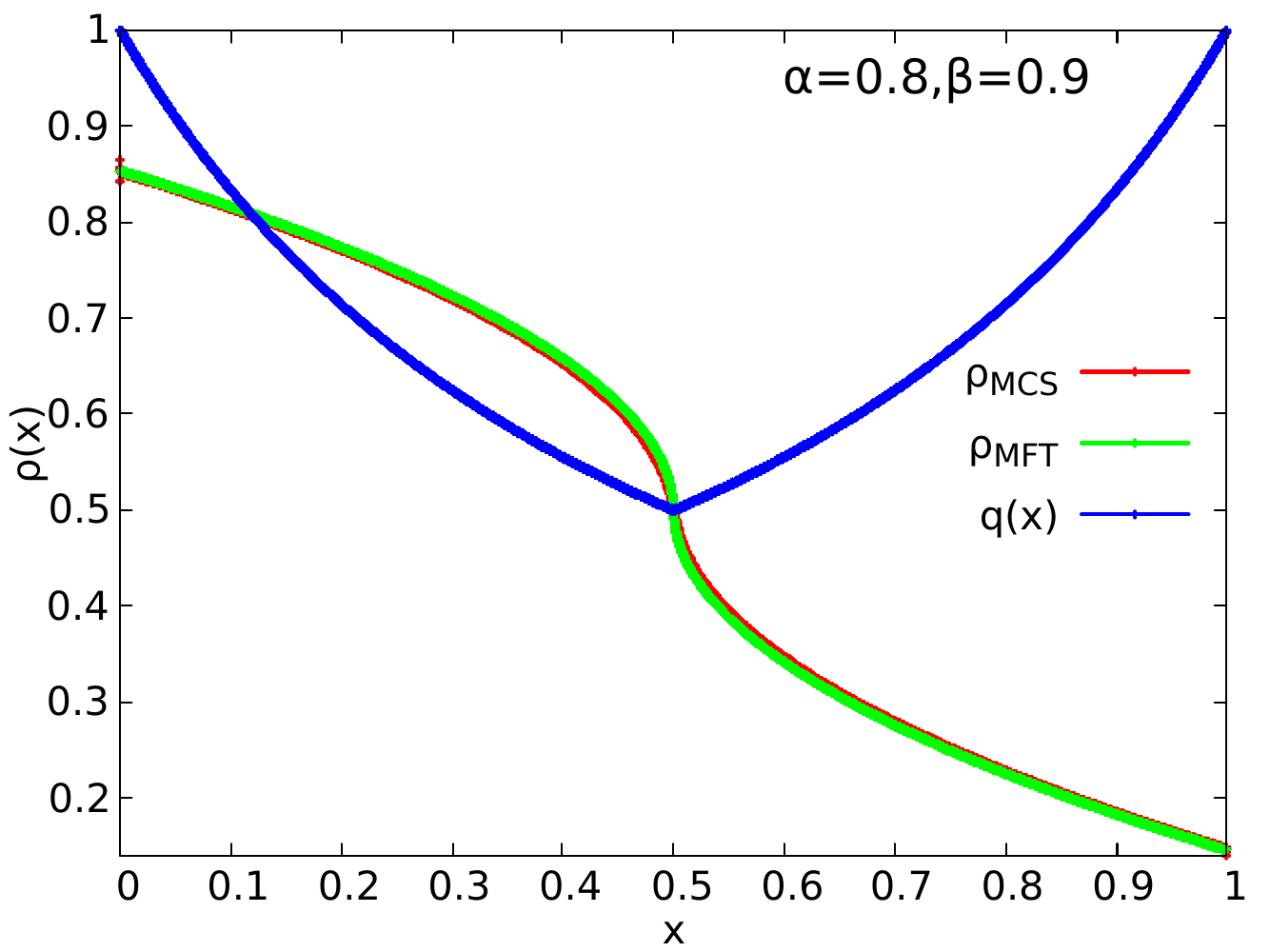}\vskip0.7cm
  \includegraphics[width=8.3cm]{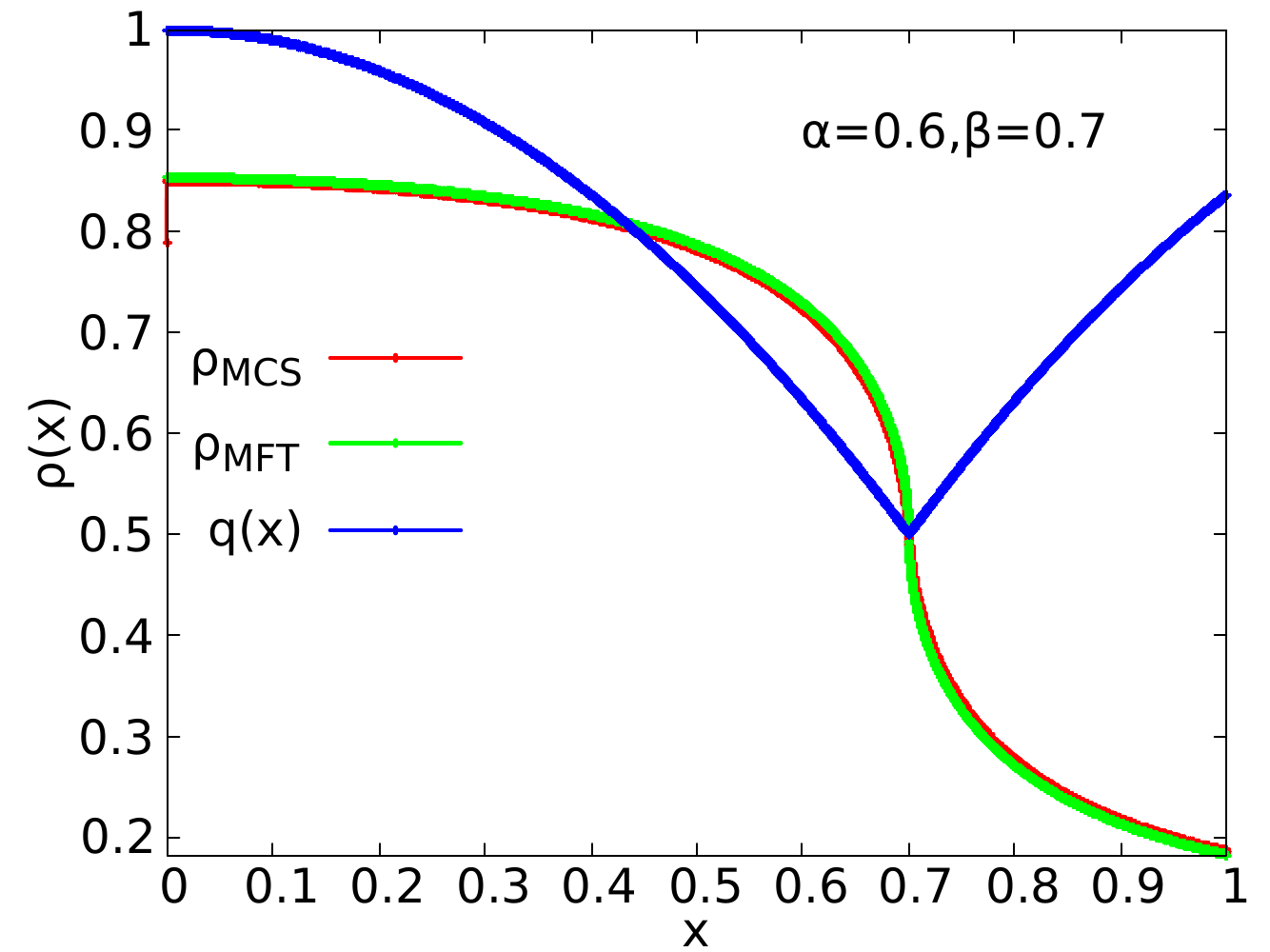}\hfill
  \includegraphics[width=8.3cm]{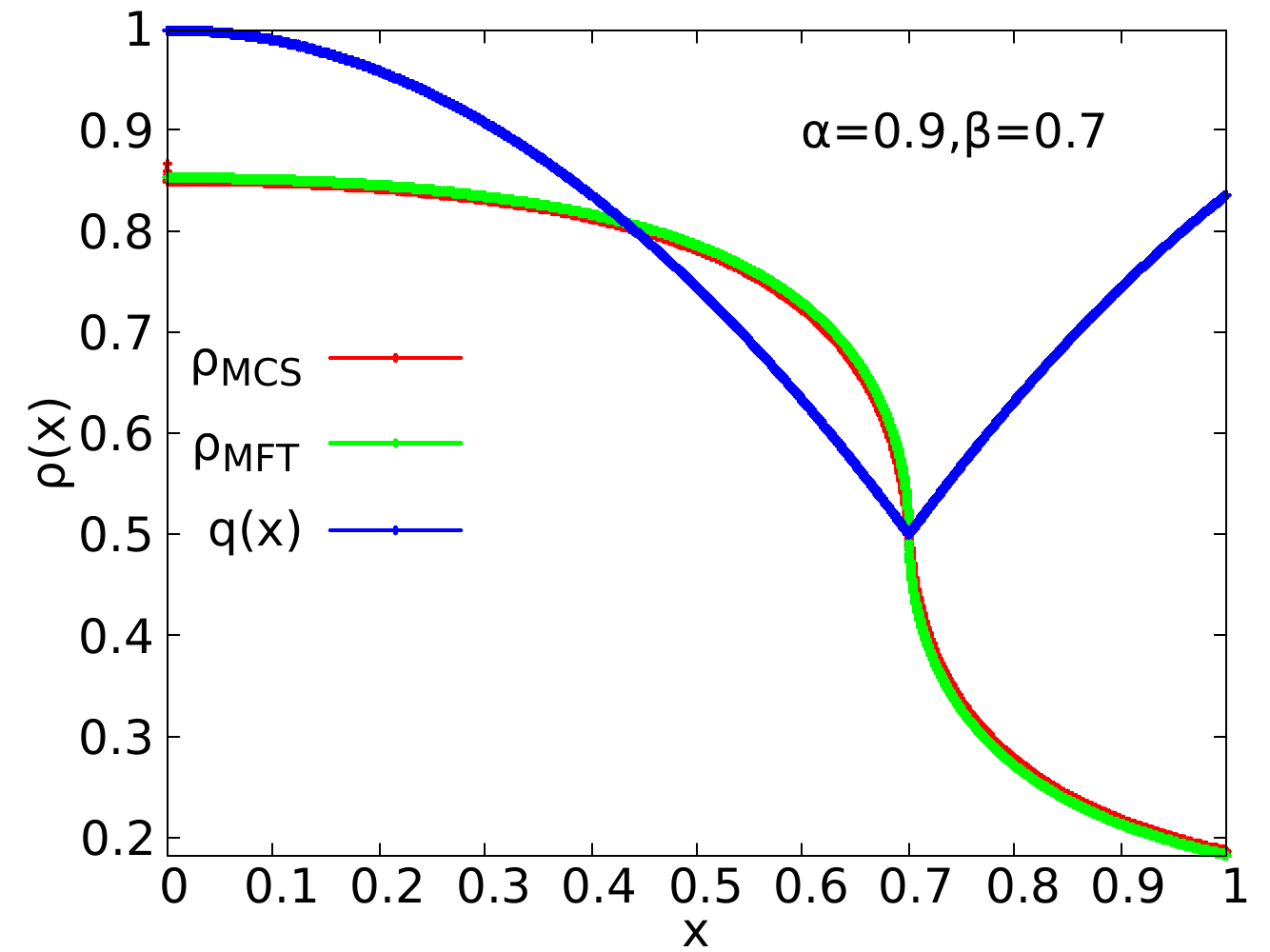}
  \caption{Plots of the steady state density $\rho(x)$ versus $x$ in the MC phase for different choices of the hopping rate functions and for different sets of values of $\alpha$ and $\beta$. (top) $q(x)$ as given in Choice I, (bottom) $q(x)$ as given in Choice II. In each plot, the green line represents the MFT prediction, the red points are from the corresponding MCS study; the blue line represents $q(x)$. Good agreement between the MFT and MCS predictions can be seen (see text).}\label{mc-plot}
 \end{figure}

 \end{widetext}

 \paragraph{MC phase:-} The steady density in the MC phase, $\rho_\text{MC}(x)$ is somewhat tricky to calculate. We already know that the steady state current in the MC phase 
 \begin{equation}
 J_\text{MC}= q_{min}/4, \label{jmc}
 \end{equation}
  which can be used either in $\rho_+(x)$ or $\rho_-(x)$, with these two solutions become identical (=1/2) at $x=x_0$, at which point $q(x_0)=q_\text{min}$. MCS studies reveal that a part of $\rho_\text{MC}(x)$ is bigger than 1/2, whereas elsewhere it is smaller than 1/2. Thus in order to construct $\rho_\text{MC}(x)$, we must use both $\rho_+(x)$ and $\rho_-(x)$, i.e., $\rho_\text{MC}(x)$ is a combination of $\rho_+(x)$ and $\rho_-(x)$, with the two segments meeting at $x_0$. { Whether $\rho_+(x)$ or $\rho_-(x)$ is to be used to construct the left or right segments (with respect to  $x_0$) remains undetermined in MFT, revealing a limitation of MFT vis-\'a-vis hydrodynamic approaches. One can of course use the MCS result as an input at this stage. However, we can instead use heuristic arguments to settle this. To proceed, we note that the model can be imagined to be composed of {\em two} inhomogeneous TASEPs $L_T$ and $R_T$ respectively on the left and right of $x_0$, which are joined at $x_0$. We can now determine the densities and phases of $L_T$ and $R_T$ separately, combining which the density profile along the full TASEP channel in its MC phase can be obtained. To proceed further, we note that at $x_0$, the hopping rate from $L_T$ to $R_T$ is $q_\text{min}$. Furthermore, the densities at the ``exit'' and ``entry'' points (both of which are nothing but $x_0$) of $L_T$ and $R_T$ are 1/2, since $\rho_+(x_0)=\rho_-(x_0)=1/2$ (see above). In addition, $L_T$ and $R_T$ have {\em no} boundary layers at $x_0$, i.e., no boundary layers at their exit and entry points, respectively. Since in the MC phase there is one boundary layer at each of $x=0$ and $x=1$, $L_T$ and $R_T$ have boundary layers at their ``entry'' (i.e., $x=0$) and ``exit'' (i.e., $x=1$), respectively. Let us now focus specifically on $L_T$. Consider first that the density profile of a uniform open TASEP at its HD-MC phase boundary is given by $\rho=1/2$ (corresponding to a steady state current of value 1/4) with a boundary layer at the entry end (i.e., $x=0$ for $L_T$) and no boundary layer at the exit end (i.e., $x=x_0$ for $L_T$).  In analogy then the density profile of the segment $L_T$ should actually resemble the density profile of a nonuniform open TASEP at its HD-MC phase boundary. This implies that $\rho(x)$ in $L_T$, i.e., for $0\leq x\leq x_0$, should be given by $\rho_+(x)$ with $J=q_\text{min}/4$. To find out the density profile in the remaining part, i.e., the density profile for $x_0\leq x\leq 1$, we now apply similar arguments on $R_T$. This gives that the density in the segment $R_T$ should actually correspond to the density in an open nonuniform TASEP at the boundary between its LD and MC phases. Hence, its density profile should be given by $\rho_-(x)$ with $J=q_\text{min}/4$, which holds for $x_0\leq x\leq 1$.  Now combining the densities in $L_T$ and $R_T$, on the whole, therefore $\rho_\text{MC}(x)$, the density profile in the MC phase of the present model, is given by $\rho_+(x)$ between 0 and $x_0$, and  $\rho_-(x)$ between $x_0$ and 1.  Alternatively,  we can directly appeal to the fact that in an open TASEP with uniform hopping in its MC phase, $\rho(0)>1/2$ and $\rho(1)<1/2$~\cite{blythe}. Since the general solution for $\rho(x)$ in the MC phase with an arbitrary $q(x)$ must reduce to the well-known solution of the density in an open uniform TASEP in its MC phase when the space-dependence of $q(x)$ gets progressively weaker as $q(x)$ approaches a constant, in order to construct $\rho_\text{MC}(x)$ in our case, we should use $\rho_+(x)$ to the left of $x_0$ and $\rho_-(x)$ to its right. Both the above heuristic arguments agree well with our MCS results; see Fig.~\ref{mc-plot} for plots of $\rho_\text{MC}(x)$  versus $x$ for two choices of $q(x)$, each of which agrees with the MFT results. Hydrodynamic approaches such as the one developed in Ref.~\cite{erdmann} should provide a more formal basis to our above heuristic arguments. }  Our analysis here further implies that if $x_0$, the location of $q_\text{min}$ is not in the bulk, but at the extreme ends (i.e., $x=0,1$), $\rho_\text{MC}(x)$ will consist of only $\rho_-(x)$ or $\rho_+(x)$. Interestingly, this means in general the average density in the MC phase (averaged over the whole TASEP) can be more or less than 1/2! This is clearly in contrast to TASEP with uniform hopping, where the average density in the MC phase is 1/2. 
  
  
 \subsection{Phase diagram}\label{phase}
 
 
 We now discuss the conditions to obtain the phase diagram and the phase boundaries in the $\alpha,\beta$-plane. First consider the boundary between the LD and HD phases. In the LD phase, the bulk current is given by $J_\text{LD}$ in (\ref{jld}), whereas the bulk current in the HD phase is given by $J_\text{HD}$ in (\ref{jhd}). The two phases meet when $J_\text{LD}=J_\text{HD}$, which gives the phase boundary between the LD and HD phases that has the form
 \begin{equation}
  q(0)\alpha(1-\alpha)=q(1)\beta(1-\beta).
   \end{equation}
This is a quadratic equation in $\beta$ in terms of $\alpha$ with two solutions $\beta_\pm$:
\begin{equation}
 \beta_\pm=\frac{1}{2}\left[1\pm \sqrt{1-4\mu\alpha(1-\alpha)}\right], \label{ld-hd-boundary}
\end{equation}
where $\mu\equiv  q(0)/q(1)$, which can be bigger or smaller than unity.   Since $\beta<1/2$ for the HD phase, $\beta=\beta_-$ gives the LD-HD phase boundary.  Phase boundary (\ref{ld-hd-boundary}) automatically reduces to $\alpha = \beta$, the well-known result for the phase boundary or the coexistence line between the LD and HD phases in an open TASEP with a uniform $q(x)$. In fact, {\em even with nonuniform} $q(x)$,  $\alpha = \beta$ {\em is the phase boundary}, so long as $q(0)=q(1)$ is maintained,  independent of the actual profile of $q(x)$.

The steady state density profile at the LD-HD coexistence line has a special structure. In an open TASEP with uniform hopping rate, it occurs on the line $\alpha=\beta<1/2$ in the $\alpha-\beta$ plane, and is actually a {\em delocalised domain wall} (DDW), which is a domain wall or a density ``shock'' whose position is not fixed but fluctuates along the whole length of the TASEP length. Moreover, the position of the domain wall is equally likely to be anywhere in the TASEP. This means the long time average of the density profile, that essentially captures the envelope of the DDW, is an inclined straight line connecting $\rho_\text{LD}=\alpha$ at the entry end and $\rho_\text{HD}=1-\beta$ at the exit end. Since the MFT neglects all fluctuations, it cannot capture this DDW.  We numerically investigate the analogue of a DDW in a uniform open TASEP in the present problem for  $q(x)$ in Choice I, which is symmetric about the mid-point $x=1/2$, and in Choice II, which is {\em not} symmetric about the mid-point, as given above~(\ref{choice1}) and (\ref{choice2}), respectively. For Choice I
 the coexistence line in the $\alpha-\beta$ plane is still given by $\alpha=\beta$ due to the symmetry of the function $q(x)$ chosen. In contrast, for Choice II there is no such symmetry, and hence the LD-HD phase boundary in the $\alpha-\beta$ plane is given by $\beta=\beta_-$; see Eq.~(\ref{ld-hd-boundary}) above. Evidently, this is {\em not} a straight line in the $\alpha-\beta$ plane. In the absence of any localising mechanism like particle non-conservation in the bulk~\cite{erwin-lk}, or global particle number conservation~\cite{niladri1,prr}, we expect to observe a DDW. However, space-dependent hopping rates imply that the domain wall should spend statistically unequal time at different positions in the bulk of the TASEP channel, which suggests a generic curvilinear envelope of the DDW under long-time averaging.  Going beyond MFT by taking into account of fluctuations should allow us write down a Fokker-Planck equation for the instantaneous position of the density shock~\cite{erwin-epje}. Solving this equation one can in principle determine the mathematical form of the envelope, which is outside the scope of the present study. Instead, we investigate the shape of the DDW envelope by extensive MCS studies. Due to the diffusive nature of the DDW fluctuations, good statistics for the DDW envelope requires averaging over $\sim L^2$ MCS steps. Because of this, we have restricted this particular study to $L=1000$ with $q(x)$ in Choice I and Choice II. To ascertain the shape of the long-time averaged envelope of the DDW, we average over $10^9$ MCS steps. Furthermore, in order 
to resolve the instantaneous (i.e., time-dependent) structures of the DDWs, we also calculate $\rho(x)$ by averaging over short time windows of $10^4$ MCS steps. The corresponding kymographs are also obtained. To generate sufficiently smooth plots, the kymographs are indeed coarse-grained over a mesoscopic lengh $r$. This mesoscopic length $r$ used to draw a kymograph, i.e. the length over which the spatial averaging is
done, should necessarily be much smaller than the system size $L$. We have used $r = 7$ in our simulations.

\begin{figure}[htb]
  \includegraphics[width=\columnwidth]{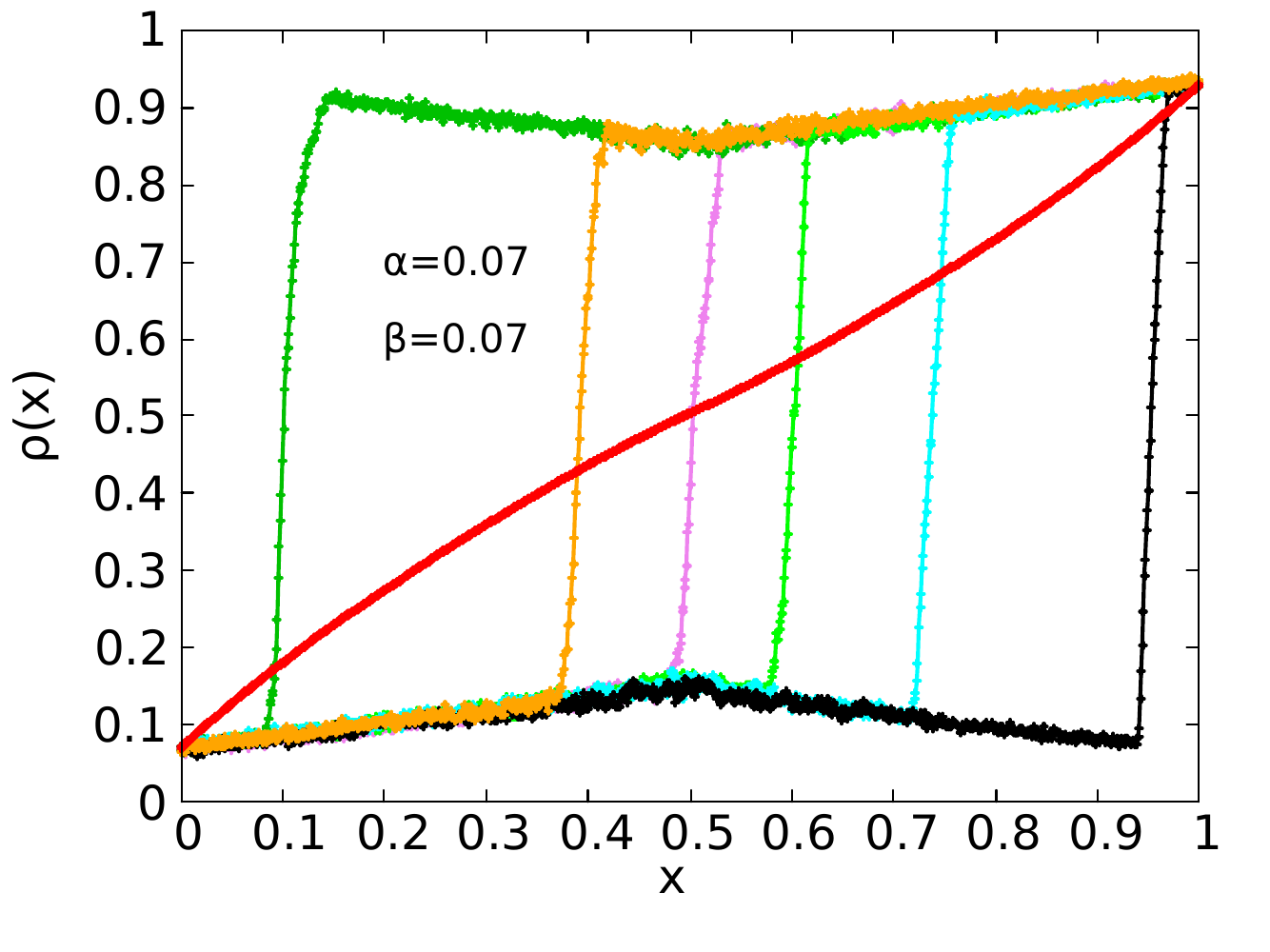}
 \caption{ Density profile $\rho(x)$ on the coexistence line with $\alpha=\beta=0.07$ with $q(x)$ as given in Choice I above. Short-time averages of $\rho(x)$ are LDWs with sharp density jumps, whose positions shift with time; the long-time average is the envelope of the moving LDWs, or a DDW, given by the inclined curved (red) line, whose precise mathematical form cannot be calculated within MFT (see text).}\label{coex-plot1}
 \end{figure}

\begin{figure}[htb]
  \includegraphics[width=\columnwidth]{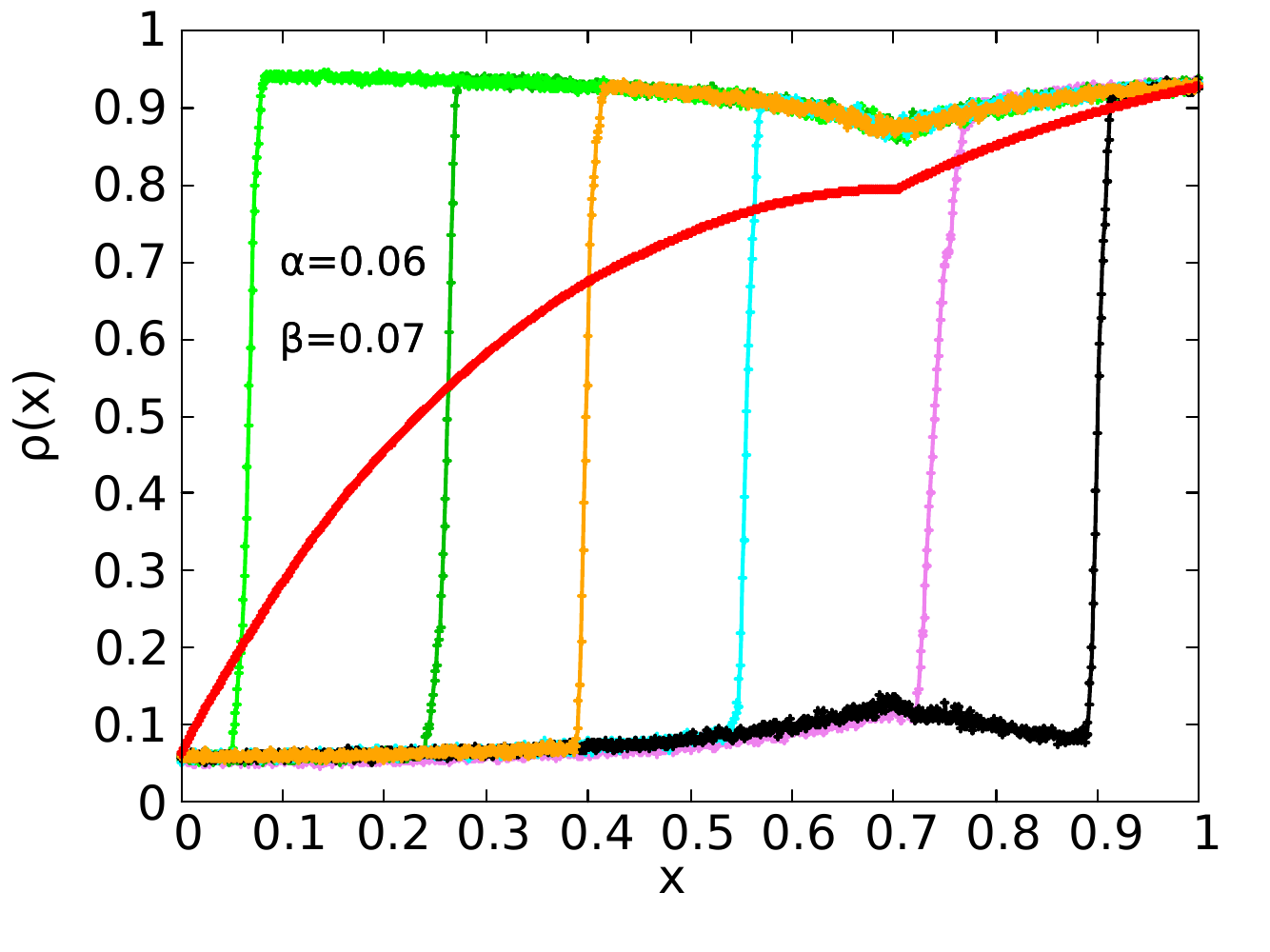}
 \caption{ Density profile $\rho(x)$ on the coexistence line with $\alpha=0.06,\,\beta=0.07$ with $q(x)$ as given in Choice II above. Short-time averages of $\rho(x)$ are LDWs with sharp density jumps, whose positions shift with time; the long-time average is the envelope of the moving LDWs, or a DDW, given by the inclined curved (red) line, whose precise mathematical form cannot be calculated within MFT (see text).}\label{coex-plot2}
 \end{figure}

\begin{figure}[htb]
  \includegraphics[width=\columnwidth]{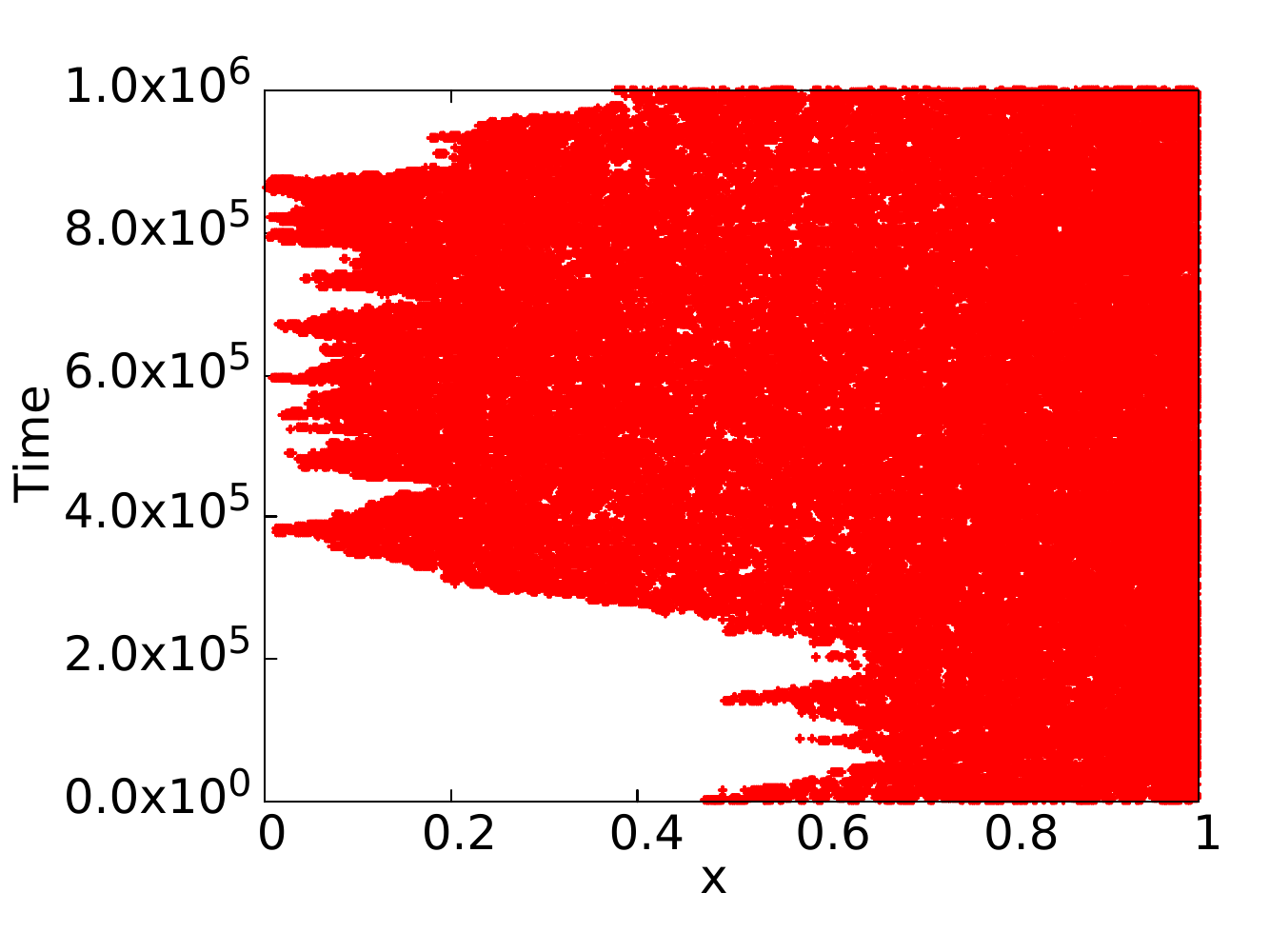}
 \caption{Kymograph for the density profile $\rho(x)$ on the coexistence line with $\alpha=\beta=0.07$ with $q(x)$ as given in Choice I above. This qualitatively reveals the delocalised nature of the domain wall.}\label{kymo1}
 \end{figure}

\begin{figure}[htb]
  \includegraphics[width=\columnwidth]{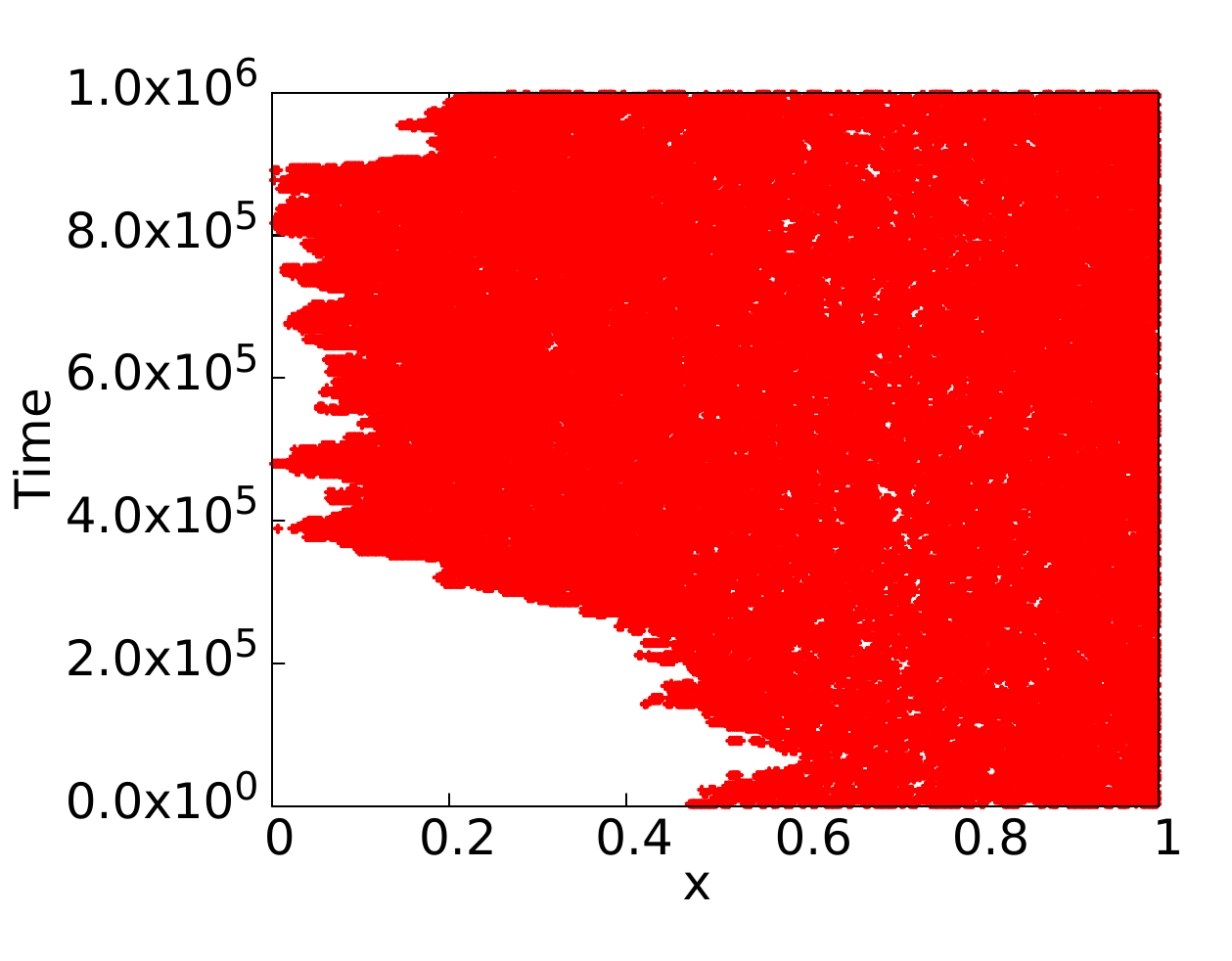}
 \caption{ Kymograph for the density profile $\rho(x)$ on the coexistence line with $\alpha=0.06,\,\beta=0.07$ with $q(x)$ as given in Choice II above. Again this qualitatively reveals the delocalised nature of the domain wall.}\label{kymo2}
 \end{figure}

We present our results on $\rho(x)$ in Fig.~\ref{coex-plot1}  and Fig.~\ref{coex-plot2}, respectively, for $q(x)$ in Choice I and Choice II. The corresponding kymographs are shown in Fig.~\ref{kymo1} and Fig.~\ref{kymo2}, respectively.

We make the following conclusions from our MCS results. First of all, both the kymographs qualitatively reveal that the domain walls are delocalised for both the choices of $q(x)$. In order to make quantitative understanding of the DDWs, we now consider the density profile plots as shown in Fig.~\ref{coex-plot1}  and Fig.~\ref{coex-plot2}. Unsurprisingly, the short-time averages in both Fig.~\ref{coex-plot1}  and Fig.~\ref{coex-plot2} reveal sharp, discontinuous structures of $\rho(x)$, which are reminiscent of a {\em localised domain wall} (LDW) in heterogeneous TASEPs in ring geometries~\cite{prr,niladri1},  for both choices of $q(x)$. The corresponding long-time averages as expected are nonlinear functions of position $x$. Nonetheless, these long time averages reveal important distinctions between these two choices of $q(x)$. For instance with Choice I, the DDW envelope shows weak departure from an inclined straight line. In addition, careful observation  reveals that the envelope takes a distinct shape close to $x=1/2$, which is the location of $q_\text{min}$, than away from it. In contrast for Choice II, $\rho(x)$ shows a strong nonlinear behaviour with $x$, and again shows a particular structure at $x\approx 0.7$, the location of $q_\text{min}$, distinct from elsewhere.  Lastly, the alert reader may notice that a general consequence of the nonlinear $x$-dependence of the DDW envelopes is that the short-time averages of the densities, though display sharp discontinuities, do not have constant densities on either side of the discontinuities; see Fig.~\ref{coex-plot1}  and Fig.~\ref{coex-plot2}. Naturally, these short-time averaged density profiles are {\em intrinsically different} from the standard heaviside step functions. At a qualitative level, we can make the following conclusion. Given that the domain wall should execute random walk along the TASEP lane, the form of the domain wall envelope for an asymmetric $q(x)$ strongly suggests that the domain wall spends different amount of time in different regions of the nonuniform TASEP lane, mimicking a random walk in a potential with a complex shape. This may be quantitatively analysed further by calculating the general profile of the DDW envelope including its shape near the minimum of $q(x)$ analytically by going beyond MFT approaches for a given $q(x)$~\cite{prr,erwin-epje}. This will be discussed elsewhere.

\begin{figure}[htb]
\includegraphics[width=1.1\columnwidth]{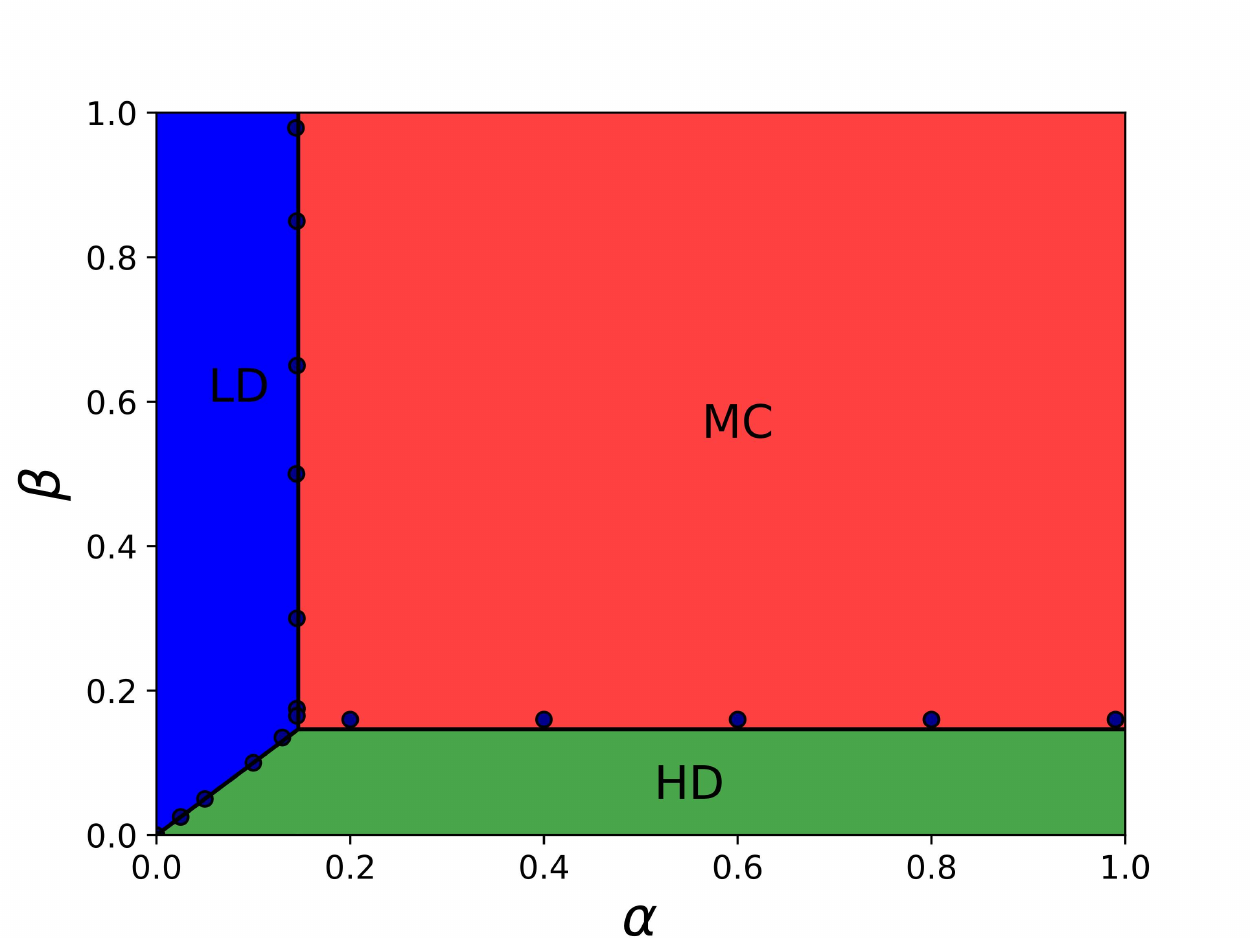}\vskip0.5cm
\includegraphics[width=1.1\columnwidth]{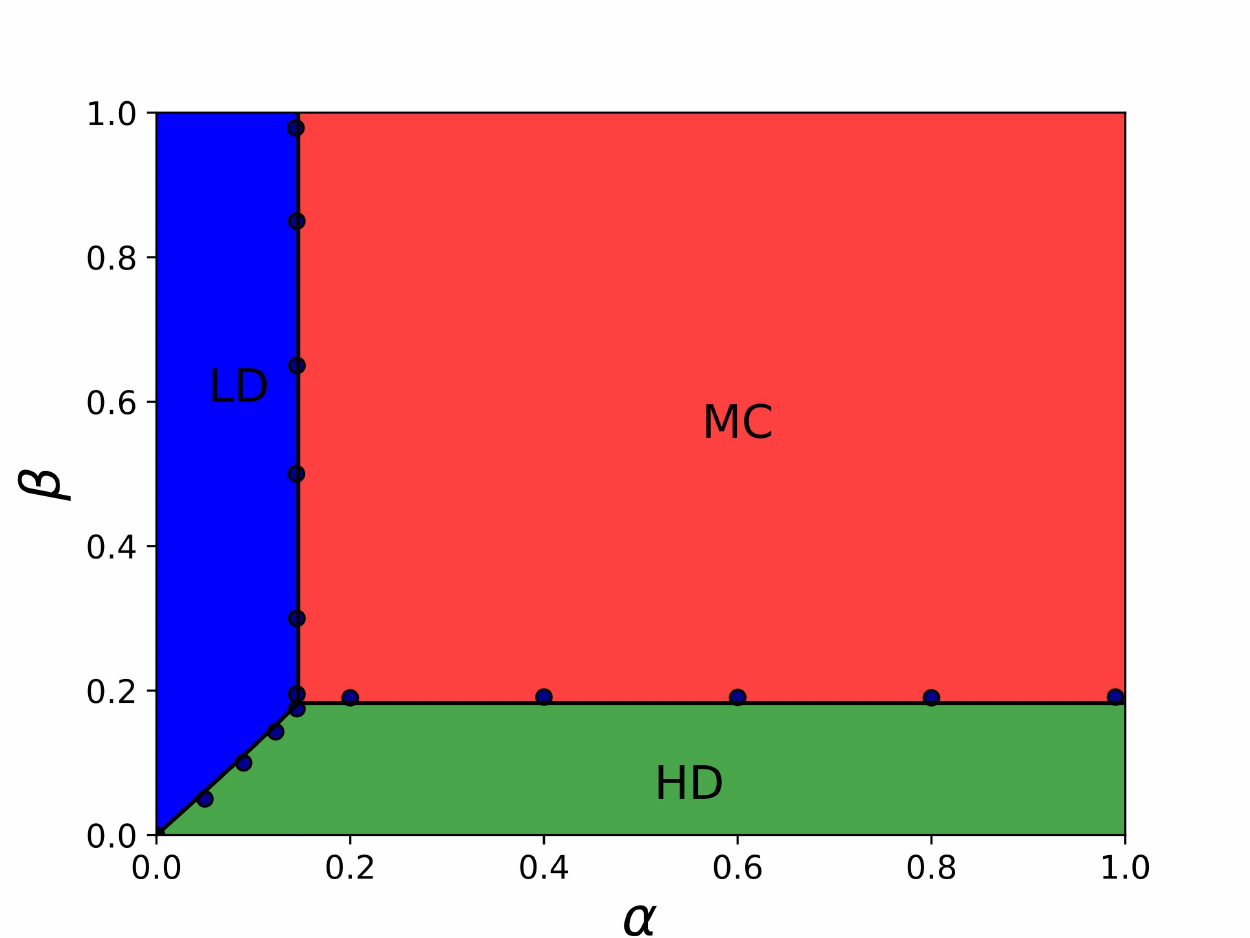}
 \caption{Phase diagram in the $\alpha-\beta$ plane with $q(x)$ in Choice I above (top), and $q(x)$ in Choice II above (bottom). Continuous lines represent the MFT predictions; discrete points are from the corresponding MCS studies, which agree well with the MFT results. LD, HD and MC phases are marked. The two phase diagrams clearly have the same topology (see text).}\label{phase1}
\end{figure}

Similar considerations allow us to obtain the LD-MC and HD-MC phase boundaries. For example, the LD-MC phase boundary is given by the condition $J_\text{LD}=J_\text{MC}$, which gives
\begin{equation}
 \alpha=\frac{1}{2}\left[1- \sqrt{1-\frac{q_\text{min}}{q(0)}}\right],\label{ld-mc-boundary}
\end{equation}
 since $\alpha<1/2$ for the LD phase. Assuming $q(i)=q_\text{min}$ for some $i$ in the bulk, the effect of a nonuniform $q(x)$ is to shift the boundary line (\ref{ld-mc-boundary}) {\em towards} the $\beta$-axis.   
Likewise, the HD-MC phase boundary is given by the condition $J_\text{HD}=J_\text{MC}$, giving
\begin{equation}
 \beta=\frac{1}{2}\left[1- \sqrt{1-\frac{q_\text{min}}{q(1)}}\right].\label{hd-mc-boundary}
\end{equation}
 since $\beta<1/2$ for the HD phase. Again with $q(i)=q_\text{min}$ for some $i$ in the bulk, the effect of a nonuniform $q(x)$ is to shift the boundary line (\ref{hd-mc-boundary}) {\em towards} the $\alpha$-axis.
The three phase boundaries meet at $\left(\left[1-\sqrt{1 - q_\text{min}/q(0)}\right]/2, \left[1-\sqrt{1 - q_\text{min}/q(1)}\right]/2\right)$.  Since $q(0),\,q(1)\geq q_\text{min}$, the general effect of a nonuniform hopping rate appears to be to enlarge the MC phase region and shrink the LD and HD phase regions in the $\alpha-\beta$-plane. Furthermore, since $q(0)\neq q(1)$ in general, the phase diagram could be asymmetric under interchange of $\alpha$ and $\beta$. Phase diagrams for $q(x)$ in Choice I and $q(x)$ in Choice II are shown in Fig.~\ref{phase1} (top) and Fig.~\ref{phase1} (bottom), respectively. Phase boundaries (\ref{ld-hd-boundary}), (\ref{ld-mc-boundary}) and (\ref{hd-mc-boundary}) calculated from MFT, and the corresponding results from MCS studies are superposed. Good agreement between the two are found. 

 Let us now make some general observations on the phase diagrams in Fig.~\ref{phase1}. Clearly, the phase diagrams in Fig.~\ref{phase1} are quantitatively different from the well-known phase diagram of an open TASEP with uniform hopping. First of all, the MC region of the phase diagrams is now distinctly bigger with space-dependent $q(x)$ than in the corresponding phase diagram with a constant hopping rate. Secondly, between the two phase diagrams presented in Fig.~\ref{phase1}, the one with $q(x)$ as given in Choice I above with $q(0)=q(1)$ [Fig.~\ref{phase1} (top)] remains unchanged under the interchange of $\alpha$ and $\beta$, same as for the phase diagram for an open TASEP with uniform hopping rate. In contrast, the phase diagram with $q(x)$ as given in Choice II, such that $q(0)\neq q(1)$, [Fig.~\ref{phase1} (bottom)] has no such symmetry under the interchange of $\alpha$ and $\beta$. These properties are consistent with our discussions above; see Eqs.~(\ref{ld-hd-boundary}), (\ref{ld-mc-boundary}) and (\ref{hd-mc-boundary}).  This firmly establishes the connections between the quantitative forms of the phase diagrams with the different choices of the hopping rate functions, a key quantitative outcome from the present study.  Nonetheless, the phase diagrams above have the same topology as that for an open TASEP with uniform hopping: all of them have three phases, which meet at a common point, establishing a degree of universality in the phase diagrams that complements the results of Ref.~\cite{prr}.


\section{Phase transitions}\label{phase-trans}

\begin{widetext}

\begin{figure}[htb]
 \includegraphics[width=5.9cm]{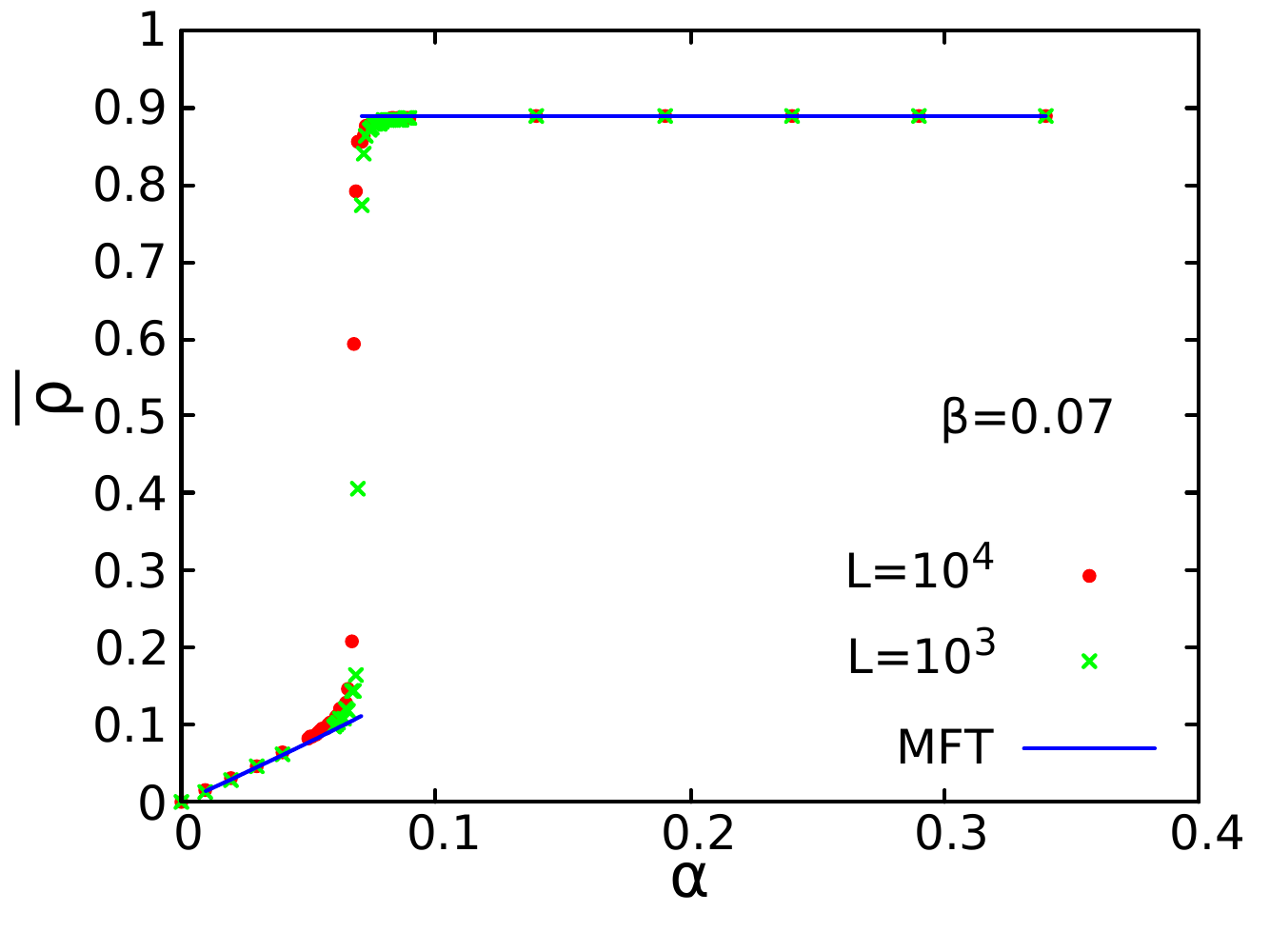}\hfill
  \includegraphics[width=5.9cm]{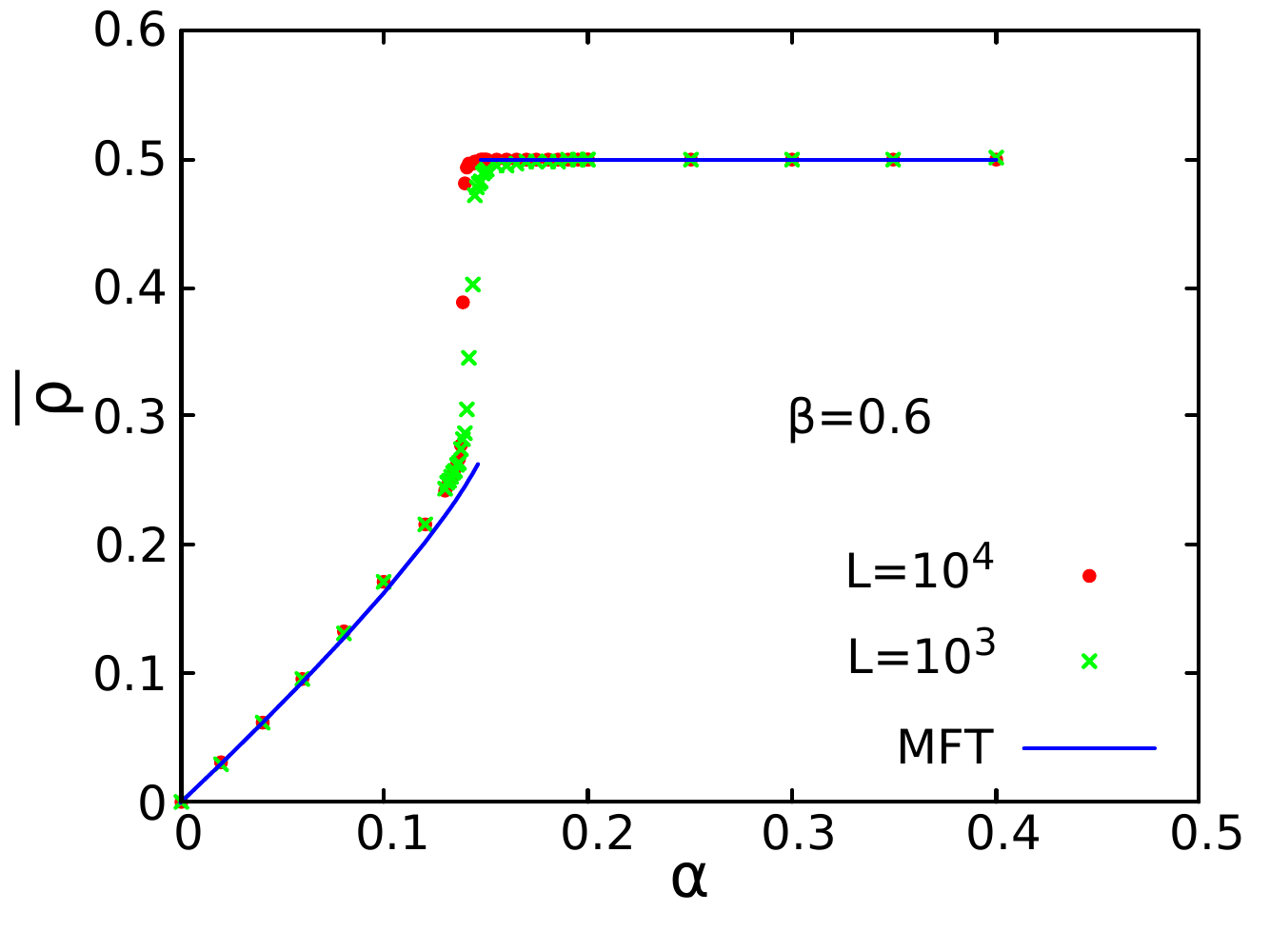}\hfill
 \includegraphics[width=5.9cm]{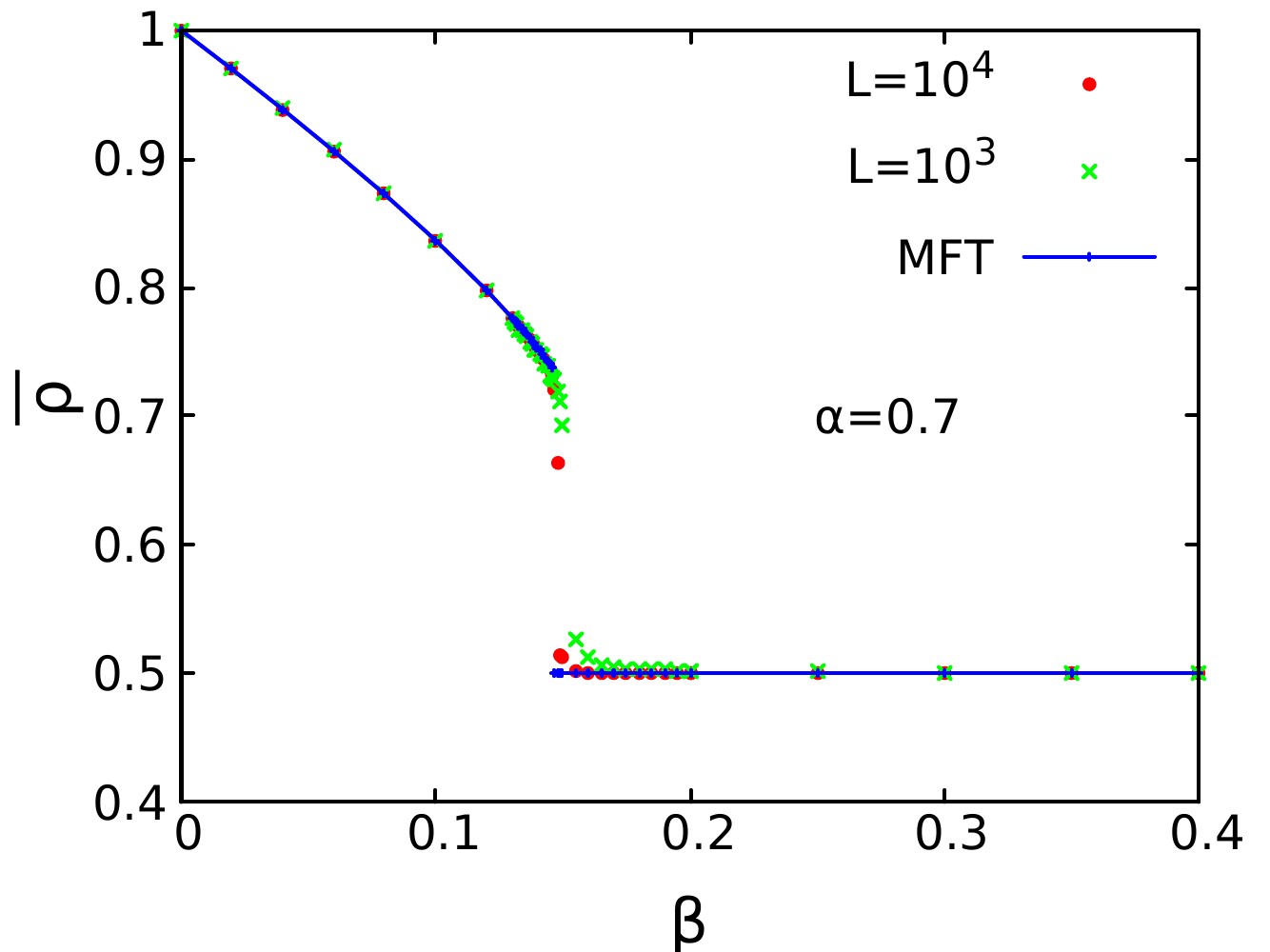}
 \caption{Plots showing phase transitions with $q(x)$ in Choice I given in (\ref{choice1}) from MFT (continuous lines) and MCS (points) studies: (left) $\overline\rho$ versus $\alpha$ for a fixed $\beta=0.07$. The horizontal line in the top gives $\overline{\rho}_\text{HD}\approx 0.9$ that is independent of $\alpha$, and the inclined line near the origin gives $\overline{\rho}_\text{LD}$ which grows with $\alpha$. (middle) $\overline\rho$ versus $\alpha$ for a fixed $\beta=0.6$. The top horizontal line gives $\overline\rho_\text{MC}\approx 0.5$ that is independent of $\alpha$ and $\beta$, and the inclined line near the origin gives $\overline{\rho}_\text{LD}$ which grows with $\alpha$. (right) $\overline\rho$ versus $\beta$ for a fixed $\alpha=0.7$. The horizontal line at the bottom gives $\overline\rho_\text{MC}\approx 0.5$ that is independent of both $\alpha$ and $\beta$, and the inclined line at the top gives $\overline\rho_\text{HD}$ that is independent of $\alpha$. All the transitions appear to be discontinuous.}\label{phase-trans-new}
\end{figure}

\end{widetext}

The original TASEP model with open boundaries and uniform hopping, the transition between the LD and HD phases are first order transitions, whereas those between the MC and LD or HD phases are second order transitions. The difference in the average bulk densities of the two phases serves as the order parameter in each of these transitions. In order to study the phase transitions in the present model, we first need to define the order parameter appropriately.
To start with, we define the mean density
\begin{equation}
 \overline \rho_a\equiv \frac{1}{L}\int_0^1 \rho_a(x) \,dx,\label{op-defn}
\end{equation}
for the phase $a$, where $a=$ LD, HD or MC phase.  Since $\rho_\text{LD}(x)<1/2$ in the bulk of the system, $\overline \rho_\text{LD}<1/2$ necessarily. Similarly, $\overline \rho_\text{HD}>1/2$ necessarily. Interestingly, $\overline \rho_\text{MC}$ need not be 1/2, in contrast to conventional open TASEPs with uniform hopping.  In fact, in the present study, with $q(x)$ in Choice I above, $\overline \rho_\text{MC}\approx1/2$~\cite{comm1} due to the symmetry of $q(x)$ and hence $\rho_\text{MC}$ about $x=1/2$. In contrast, for $q(x)$ in Choice II above, $\overline\rho_\text{MC}>1/2$. { Order parameter (\ref{op-defn}) clearly generalises the order parameter in a uniform TASEP, which is just the bulk density.} With this, considering the mean density as the order parameter, the transition between the LD and HD phases is a first order transition with 
\begin{equation}
\mathbb{O}_\text{HD-LD}\equiv \overline \rho_\text{HD}-\overline \rho_\text{LD} \label{op-ld-hd}
\end{equation}
showing  a jump across the LD-HD phase boundary. This jump, given by the magnitude of $\mathbb{O}_\text{HD-LD}$ is to be calculated on the phase boundary between the LD and HD phases, and clearly depends on $\alpha$ (or equivalently $\beta$). The finite jump of $\mathbb{O}_\text{HD-LD}$ tells us that the phase transition in question is a first order transition. To study the phase transitions between the MC and LD or HD phases, we similarly consider 
\begin{eqnarray}
 \mathbb{O}_\text{LD-MC}&\equiv& \overline \rho_\text{LD}-\overline \rho_\text{MC}, \label{op-ld-mc}\\
 \mathbb{O}_\text{HD-MC}&\equiv& \overline \rho_\text{HD}-\overline \rho_\text{MC}. \label{op-hd-mc}
\end{eqnarray}
It is easy to see that these order parameters (\ref{op-ld-hd})-(\ref{op-hd-mc}) reduce to the respective bulk density differences in an open TASEP with uniform hopping rates.
{ The transitions would be second order if the respective order parameters defined above would vanish at the corresponding phase boundaries. Else, if instead they show discontinuities, the transitions are first order in nature. In this Section, we focus on the phase transitions in the restricted case with symmetric $q(x)$ with a single minimum (as above). The more general cases including asymmetric $q(x)$ will be discussed elsewhere in future. We use both MFT and MCS to analyse the phase transitions. We first study the LD-HD transition. To that end, we consider Eq.~(\ref{rho+}) or Eq.~(\ref{rhohd}) for $\rho_\text{HD}(x)$, and correspondingly Eq.~(\ref{rho-}) or Eq.~(\ref{rhold}) for $\rho_\text{LD}(x)$. Since both $J_\text{LD},\,J_\text{HD}<J_\text{max}=J_\text{MC}=q_\text{min}/4$, we conclude that the discriminant in each of Eq.~(\ref{rho+}) or Eq.~(\ref{rhohd}) and Eq.~(\ref{rho-}) or Eq.~(\ref{rhold}), giving $\rho_\text{HD}(x)>1/2$ and $\rho_\text{LD}(x)<1/2$ at {\em all} $x$. This in turn implies $\overline\rho_\text{HD}>1/2$ and $\overline \rho_\text{LD}<1/2$ necessarily. This holds even at the transition point making $\overline\rho_\text{HD} \neq \overline \rho_\text{LD}$, which in turn means $\mathbb{O}_\text{HD-LD}$ is {\em finite} at the transition. Thus the LD-HD transition in the MFT is first order, as in an open TASEP with constant hopping rates. We now turn to the LD-MC transition, which is a second order transition in an open TASEP with constant hopping. To analyse this for symmetric $q(x)$, we use the fact that $\overline \rho_\text{MC}=1/2$ (see above), and again consider Eq.~(\ref{rho-}) or Eq.~(\ref{rhold}) for $\rho_\text{LD}(x)$. At the LD-MC transition $J_\text{LD}=J_\text{MC}=q_\text{min}/4$. Since $q(x)\geq q_\text{min}$ at any $x$, the discriminant in Eq.~(\ref{rho-}) or Eq.~(\ref{rhold}) is generally positive except at discrete points where $q(x)=q_\text{min}$, which holds even at the LD-MC transition. Thus $\rho_\text{LD}(x)$ and hence $\overline \rho_\text{LD}<1/2 = \overline \rho_\text{MC}$ at the transition point. Surprisingly, this means $\mathbb{O}_\text{LD-MC}$ {\em does not} vanish at the transition, giving a {\em first order transition} here, in contrast to a second order LD-MC transition in an open TASEP with uniform hopping. Similar arguments can be used to show that in MFT, the HD-MC transition is also a first order transition, again in contrast to an open TASEP with uniform hopping. Thus, rather unexpectedly MFT predicts that {\em all the three} transitions are first order with a symmetric $q(x)$ having one minimum.  To verify these MFT predictions for our model numerically, we have studied the nature of the phase transitions numerically across the LD-HD, LD-MC and HD-MC phase boundaries. More specifically, we calculate: 

(i) $\overline \rho$ as a function of $\alpha$ for a fixed $\beta=0.07$, as $\alpha$ approaches the LD-HD phase boundary; see Fig.~\ref{phase-trans-new}(left) for a plot of the average density as a function of $\alpha$. On one side of the transition, $\overline\rho$ is the mean LD phase density $\overline \rho_\text{LD}$ that rises with $\alpha$; on the other side of it, $\overline\rho$ is $\overline\rho_\text{HD}$ that remains independent of $\alpha$ for a fixed $\beta$. This plot clearly shows a jump in $\overline\rho$ across the transition, meaning a first order transition, akin to an open TASEP with uniform hopping.   

(ii)  $\overline \rho$ as a function of $\alpha$ for a fixed $\beta=0.6$, as $\alpha$ approaches the LD-MC phase boundary; see Fig.~\ref{phase-trans-new}(middle) for a plot of the average density as a function of $\alpha$. On one side of the transition, $\overline\rho$ is the mean LD phase density $\overline \rho_\text{LD}$ that rises with $\alpha$; on the other side of it, $\overline\rho$ is $\overline\rho_\text{MC}$ that remains independent of $\alpha$ as well as $\beta$. Surprisingly, this plot clearly shows a jump in $\overline\rho$ across the transition, meaning a {\em first order transition},  in contrast to an open TASEP with uniform hopping.  

(iii)  $\overline \rho$ as a function of $\beta$ for a fixed $\alpha=0.7$, as $\beta$ approaches the HD-MC phase boundary; see Fig.~\ref{phase-trans-new}(right) for a plot of the average density as a function of $\beta$. On one side of the transition, $\overline\rho$ is the mean HD phase density $\overline \rho_\text{HD}$ that decreases as $\beta$ increases; on the other side of it, $\overline\rho$ is $\overline\rho_\text{MC}$ that remains independent of $\alpha$ as well as $\beta$. Surprisingly, this plot clearly shows a jump in $\overline\rho$ across the transition, again implying a {\em first order transition},  again in contrast to an open TASEP with uniform hopping.  

Our MCS studies show that in all these cases  there is a jump in the density at the respective phase boundary, implying a discontinuous or a first order transition, in agreement with the MFT prediction for the same. This is a truly novel result, that shows how quenched disorder can alter the order of phase transitions. To benchmark our numerical codes, in Fig.~\ref{phase-trans-open} in Appendix~\ref{app1} we have shown the analogous plots for the LD-HD, LD-MC and HD-MC transitions for an open TASEP with uniform hopping. Unsurprisingly. the plots in Fig.~\ref{phase-trans-open} show a first order LD-HD transition and second order LD-MC and HD-MC transitions, both in the MCS and MFT studies. That quenched disorder can change the order of transitions in pure models is well-known. For instance, Ref.~\cite{prb} shows that sufficiently strong quenched disorder can make the magnetic transition in ferromagnetic manganites first order. Similarly, quenched disorder can introduce a first order transition in the well-known Kuramoto model of oscillator synchronization~\cite{kura}. The current study forms yet another such example, and possibly the first of its kind in TASEP-like driven models with open boundary conditions. }

 \section{Summary and outlook} \label{summ}
 
 We have thus studied the  totally asymmetric exclusion process with open boundaries having spatially smoothly varying hopping rates. Our study reveals the universal form of the phase diagrams for generic smooth hopping rates. Our results are sufficiently general and applies to any smoothly varying hopping rate functions. We construct the mean-field theory, and use that to outline a scheme to calculate steady state density profiles. { Our method  directly gives the steady state densities in terms of the current $J$ almost immediately by using the spatial constancy of the latter in the steady states. The different phases are then analysed by varying the boundary conditions in a straightforward manner.}  Unsurprisingly, the  bulk steady state densities are {\em generically space varying}, unlike those in the conventional TASEP with uniform hopping (except along the special line $\alpha=\beta < 1/2$). These match well with those obtained from the MCS studies, lending credence to our mean-field analysis. Because of the spatially varying densities, the conventional way to characterise the phases via the densities, i.e., $\rho_\text{LD}<1/2,\,\rho_\text{HD}>1/2$ and $\rho_\text{MC}=1/2$ in the bulk of the TASEP no longer holds. Rather one needs to resort to the equivalent conditions to decide the phases, since the current $J$ is a constant. This together with the condition that $\rho_\text{LD}(x)<1/2$ and $\rho_\text{HD}(x)>1/2$ everywhere in the bulk, allows us to distinguish the LD and HD phases. Further, the maximum steady state current that the system can sustain is no longer 1/4, but is $q_\text{min}/4$, where $q_\text{min}$ is the minimum hopping rate. Surprisingly, our theory shows that the average bulk density in the MC phase can be more or less than 1/2, in direct contrast with conventional open TASEPs with uniform hopping. We show that the general effect of spatially varying hopping rates is to enlarge the MC region of the phase space, while shrinking the LD and HD regions. Furthermore, our work elucidates the universal phase diagram for various choices of the hopping rate function, highlighting the robustness of asymmetric exclusion process in an open system.  Lastly but not the least, our MFT and MCS studies clearly show that both the LD-MC and HD-MC transitions in the model are first order for any symmetric hopping rate function $q(x)$ having one minimum. This forms a truly novel outcome from this study. How this generalises to other forms of $q(x)$ is an interesting question to be investigated in the future.
 
 Our MFT scheme is sufficiently general. It applies for any $q(x)$ that is smoothly and slowly varying.  It would be interesting to extend our scheme to situations where $q(x)$ is smooth and slowly varying in general, but can have a few finite discontinuities. This will be discussed in the future.  It will also be important to study the effects of interactions in the systems~\cite{arvind1}. In addition, there are {\em in vivo} situations, where $q(x)$ is rapidly fluctuating in space~\cite{dna}, which breaks down the assumption of slowly varying $q(x)$.   How an equivalent analysis may be carried out for such a system, and to what degree the present results may be valid there are interesting questions to study.  Hydrodynamic approaches should be promising in this regards, which already provides initial clues to this problem~\cite{erdmann}; see also Ref.~\cite{astik-prr} for a comprehensive hydrodynamics-based field theory approach to this problem in a closed geometry. It would also be interesting to apply the boundary layer theory developed in Ref.~\cite{smb} on our model, and determine the stationary densities and phases. We hope our work here will provide impetus to studies along these lines in future. 
 Our results may be verified in model experiments on the collective motion of driven particles with light-induced activity~\cite{light} passing through a narrow channel. Spatial modulations of the hopping rate can be created by applying patterned or spatially varying illumination.

 \section{Acknowledgement}

We thank S. Nag and A. Basu for helpful discussions. S.M. thanks SERB (DST), India for partial financial support through the CRG scheme [file: CRG/2021/001875].

\appendix

\section{Phase transitions in an open TASEP with uniform hopping}\label{app1}

We numerically study the LD-HD, LD-MC and HD-MC transitions. As expected, our studies show that the LD-HD transition is a first order transition, whereas the LD-MC and HD-MC transitions are second order in nature. 

\begin{widetext}
 
 \begin{figure}[htb]
 \includegraphics[width=5.9cm]{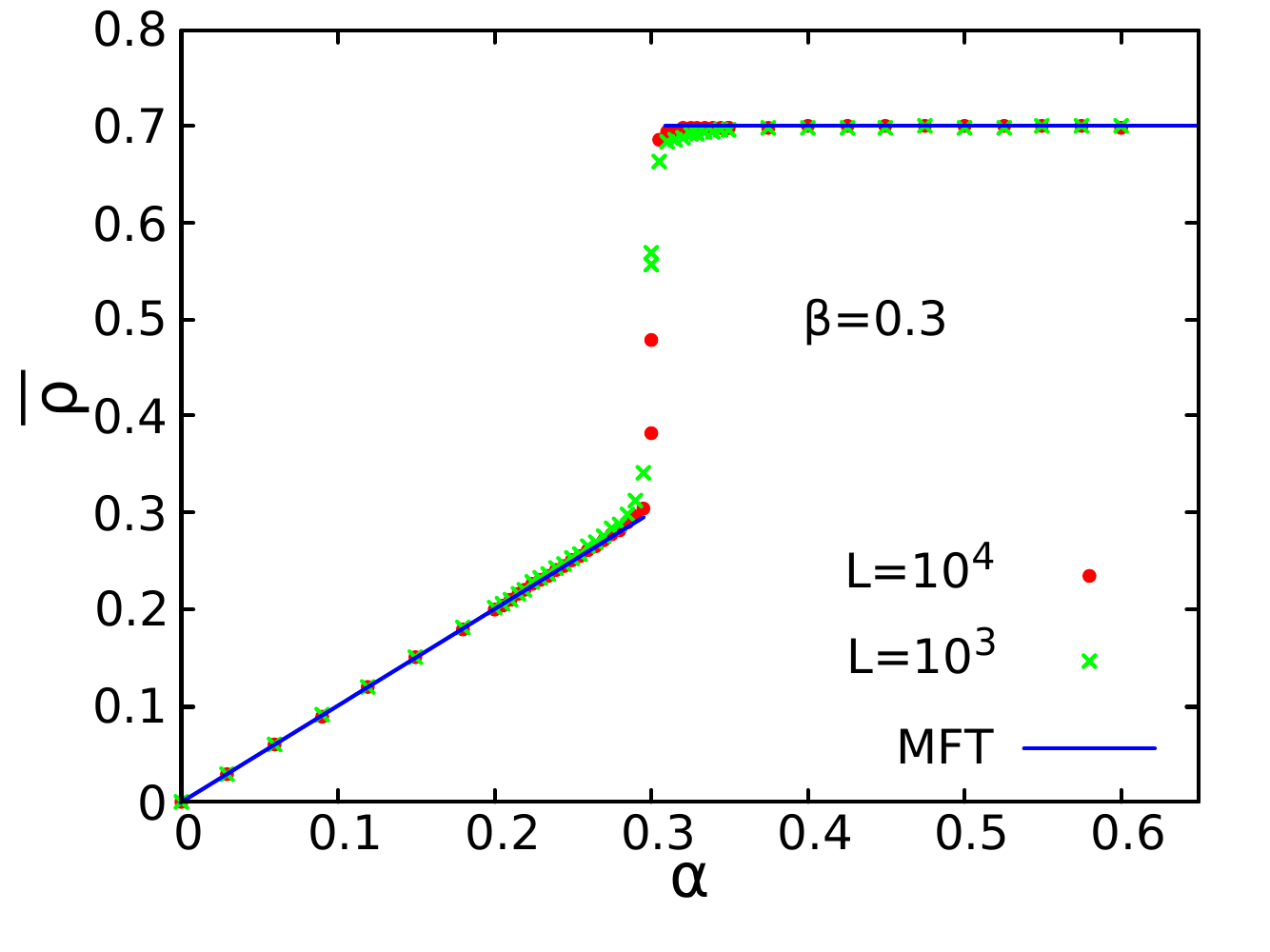}\hfill \includegraphics[width=5.9cm]{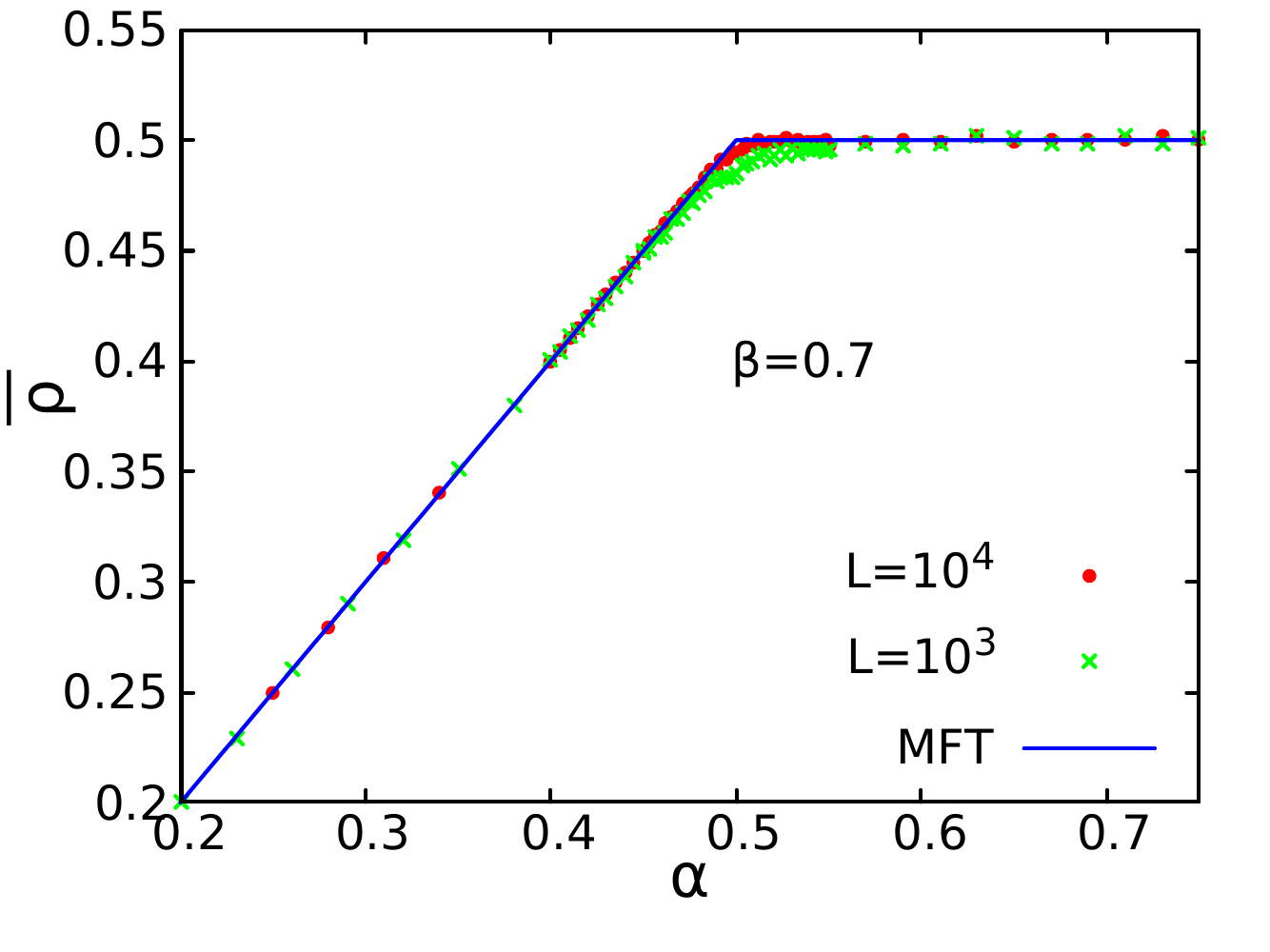}\hfill  
 \includegraphics[width=5.9cm]{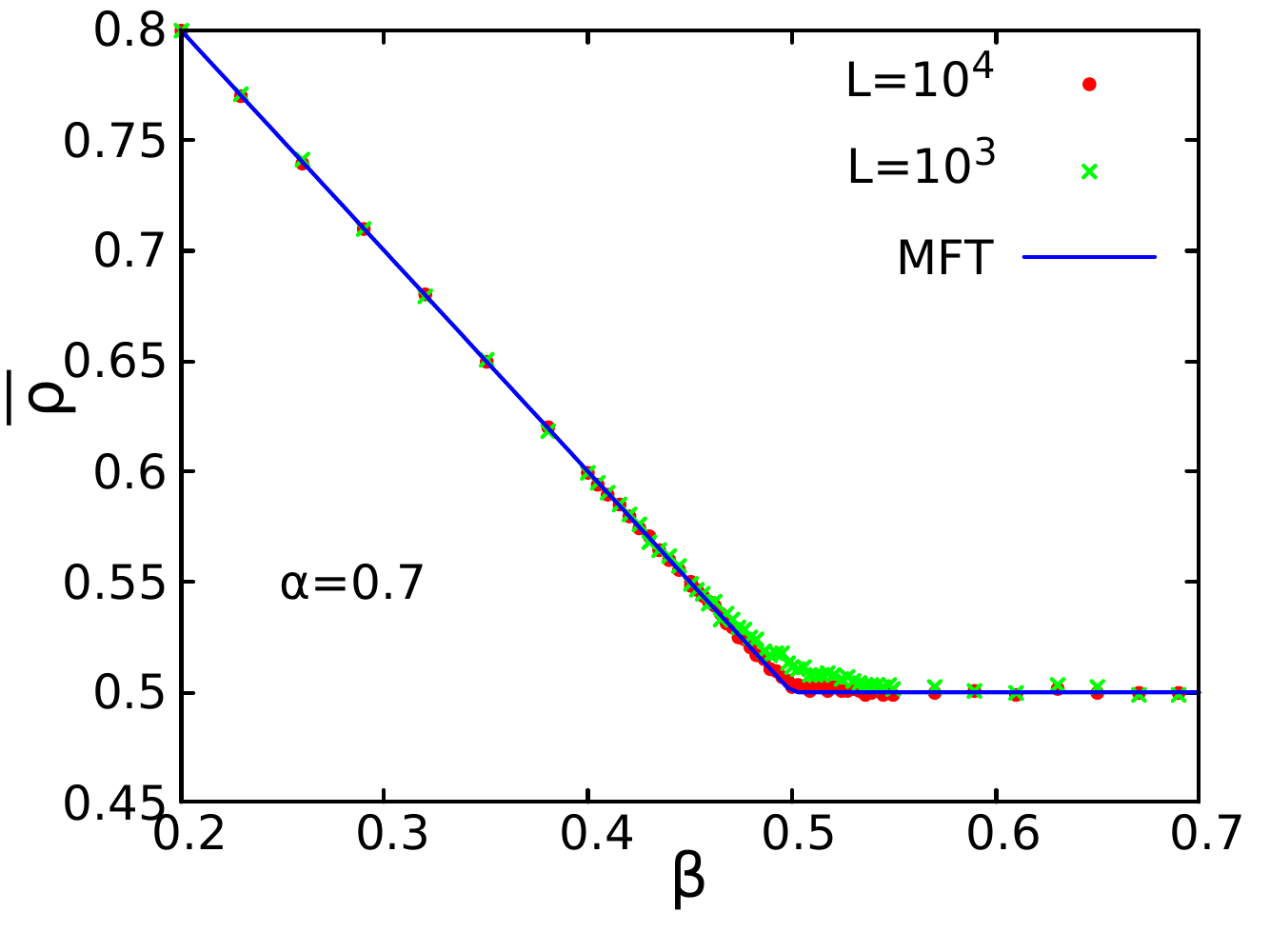}
  \caption{Plots of the densities showing phase transitions from MFT and MCS studies in an open TASEP with a constant hopping rate: (a) LD-HD phase transition, (b) LD-MC transition, and (c) HD-MC transition. Both MFT and MCS show that (a) is a first order, and (b) and (c) are second order transitions, as is well-known for open TASEPs with uniform hopping.}\label{phase-trans-open}
 \end{figure}

\end{widetext}


\begin{thebibliography}{99}
\bibitem{comm1} This would be strictly true if $\alpha=\beta$ due to the symmetry of $q(x)$. Even when $\alpha\neq \beta$ in the MC phase, $\overline \rho_\text{MC}\approx 1/2$ holds to a high accuracy, as can be seen from our MCS study results. This is presumably due to the fact the $\alpha\neq \beta$ affects for all practical purposes only the boundary layers.
\end{thebibliography}
\end{document}